\documentclass[sigconf]{acmart}
\usepackage{soul}
\usepackage{longtable}
\usepackage{xspace}
\usepackage{dblfloatfix}  
\usepackage{color}
\usepackage{amsmath}
\usepackage{mathtools} 
\usepackage{fontawesome5}
\usepackage{makecell}
\usepackage{enumitem}
\usepackage{graphicx}
\usepackage{enumitem}
\usepackage{pdfpages}
\usepackage{graphicx}
\usepackage{appendix}
\usepackage{subfigure} 
\usepackage{booktabs}
\usepackage{multirow}
\usepackage{placeins}
\usepackage{hyperref}
\usepackage{float}
\usepackage{soul}
\usepackage{xcolor}
\usepackage{tcolorbox}
\definecolor{dot1}{HTML}{7bff59}
\definecolor{dot2}{HTML}{ffed59}
\definecolor{dot3}{HTML}{ffb654}
\definecolor{lightgrey}{rgb}{0.7, 0.7, 0.7}

\newcommand{\idename}{Fusion 360\xspace} 
\usepackage{tikz}
\newcommand{\filledcircnum}[1]{%
    \tikz[baseline=(char.base)]{
        \node[shape=circle, fill=black, text=white, inner sep=1pt] (char) {#1};
    }%
}
\newcommand{\systemname}{FeedQUAC\xspace}

\AtBeginDocument{%
  }
\setcopyright{acmlicensed}
\copyrightyear{2025}
\acmYear{2025}
\acmDOI{XXXXXXX.XXXXXXX}

\acmConference[arxiv '25]{}{FeedQUAC}{Quick Unobtrusive AI-Generated Commentary}

\acmISBN{978-1-4503-XXXX-X/18/06}
\renewcommand\footnotetextcopyrightpermission[1]% removes footnote with conference information in first column

\begin{document}

\title{\systemname: Quick Unobtrusive AI-Generated Commentary}

\author{Tao Long}
\email{long@cs.columbia.edu}
\affiliation{%
  \institution{Autodesk Research}
  \city{San Francisco}
  \state{California}
  \country{USA}
}
\authornote{Also affiliated with Columbia University.}
% \author{Tao Long\textsuperscript{\dag}}
% \email{long@cs.columbia.edu}
% \affiliation{%
%   \institution{Autodesk Research}
%   \city{San Francisco}
%   \state{California}
% \thanks{\textsuperscript{\dag}Also affiliated with Columbia University.}

\author{Kendra Wannamaker}
\email{kendra.wannamaker@autodesk.com}
\affiliation{
  \institution{Autodesk Research}
  \city{Toronto}
  \state{Ontario}
  \country{Canada}
  }

\author{Jo Vermeulen}
\email{jo.vermeulen@autodesk.com}
\affiliation{
  \institution{Autodesk Research}
  \city{Toronto}
  \state{Ontario}
  \country{Canada}
  }

\author{George Fitzmaurice}
\email{george.fitzmaurice@autodesk.com}
\affiliation{
  \institution{Autodesk Research}
  \city{Toronto}
  \state{Ontario}
  \country{Canada}
  }

\author{Justin Matejka}
\email{justin.matejka@autodesk.com}
\affiliation{
  \institution{Autodesk Research}
  \city{Toronto}
  \state{Ontario}
  \country{Canada}
  }

\renewcommand{\shortauthors}{Long et al.}

%%%%%%%%%%%%%%%%%%%%%%%%%%%%%%%%%%%%%%%%%%%%%%%%%%%%%%%%%%%%%%
%%%%%%%%%%%%%%%%%%%%%%%%%% ABSTRACT %%%%%%%%%%%%%%%%%%%%%%%%%%
%%%%%%%%%%%%%%%%%%%%%%%%%%%%%%%%%%%%%%%%%%%%%%%%%%%%%%%%%%%%%%

\begin{abstract}
Design thrives on feedback. However, gathering constant feedback throughout the design process can be labor-intensive and disruptive. We explore how AI can bridge this gap by providing effortless, ambient feedback. We introduce \systemname, a design companion that delivers real-time AI-generated commentary from a variety of perspectives through different personas. A design probe study with eight participants highlights how designers can leverage quick yet ambient AI feedback to enhance their creative workflows. Participants highlight benefits such as convenience, playfulness, confidence boost, and inspiration from this lightweight feedback agent, while suggesting additional features, like chat interaction and context curation.  
We discuss the role of AI feedback, its strengths and limitations, and how to integrate it into existing design workflows while balancing user involvement. Our findings also suggest that ambient interaction is a valuable consideration for both the design and evaluation of future creativity support systems.

\end{abstract}

\begin{teaserfigure}
 \vspace{-5px}
 \includegraphics[width=\linewidth]{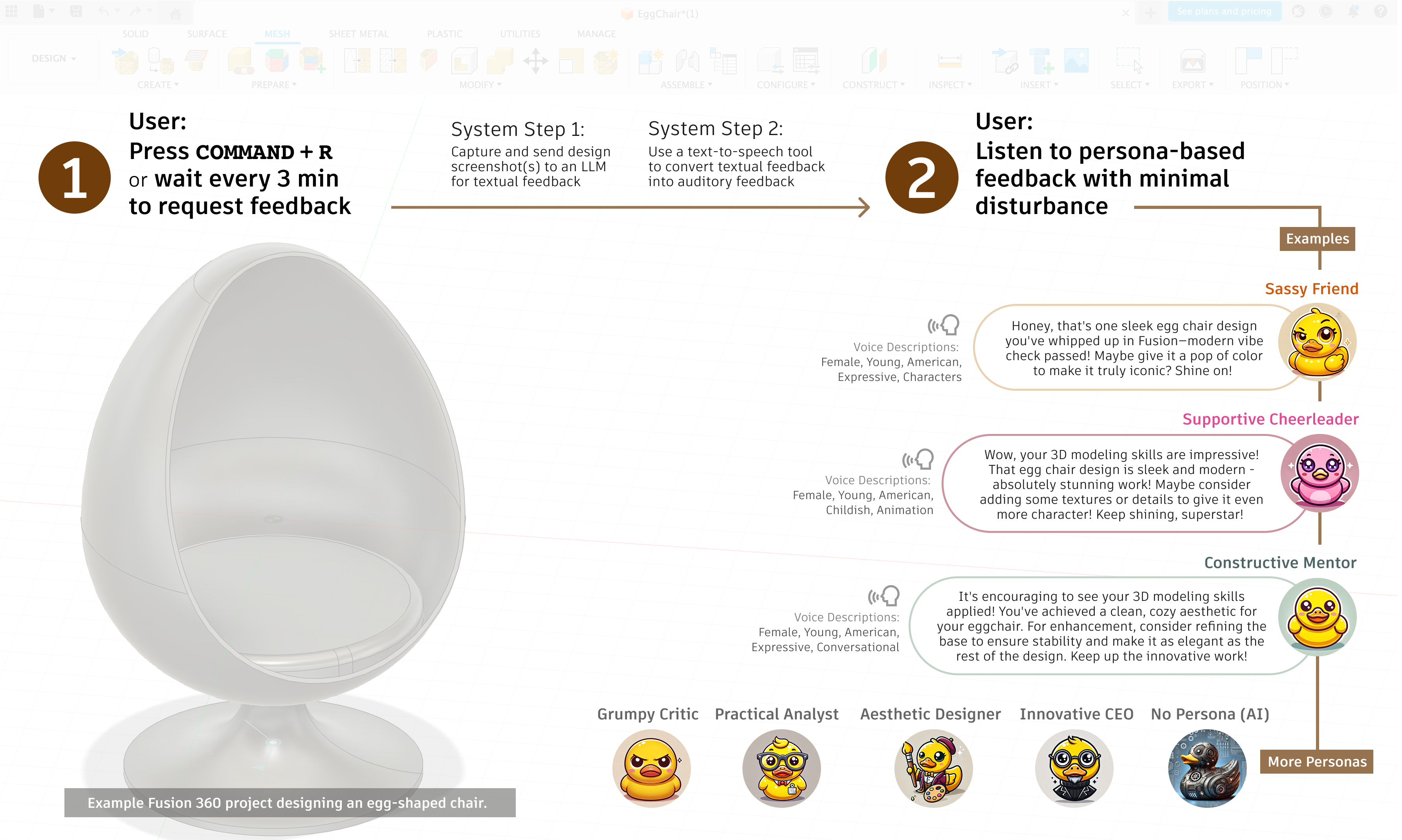}
  \caption{\systemname is an ambient design companion tool that supports designers by providing real-time, read-aloud, AI-generated feedback from diverse personas throughout the design process. With its seamless integration---offering feedback via shortcut keys, minimal screen space, and auditory cues---\systemname ensures a quick, unobtrusive feedback experience that enhances the design without interrupting the designer's workflow.}
  \label{fig:teaser}
  \Description[A teaser photo for our system, FeedQUAC.]{A teaser photo for our system, FeedQUAC. The photo includes a screenshot of the usage scenario of FeedQUAC inside a design editor, along with a simple flowchart illustrating how FeedQUAC works. }
  \vspace{12px}
\end{teaserfigure}


\maketitle
\newpage

\section{Introduction}
% \tao{new intro --- feel free to change <3}
% \\
{Feedback is important for better understanding and improving designs~\cite{feedbackImportant}. It supports the iterative nature of the design process, enabling designers to rethink and refine their vision~\cite{krishna2022setting, andradeFormativeAssessmentArts2019, anseel2009reflection, yen2017listen}. 
A recent study by E et al.~\cite{janeEpaper} explores the trade-offs in the timing of design feedback and emphasizes the value of short-form, real-time feedback throughout the design process. Their findings suggest that providing designers with quick, targeted, and frequent feedback helps them make greater improvements in real time, ultimately leading to fewer issues in the final design~\cite{janeEpaper}. 
Krishna Kumaran et al.~\cite{krishna2022setting} further demonstrate that while feedback strategies vary by expertise and design stage, the need for timely, targeted feedback remains consistent throughout the creative process.
%These short commentaries, usually consisting of 2-4 sentences, highlight and discuss the specific design changes that users made and offer validation and actionable advice to help the designer maintain momentum. After receiving instant feedback during the design process, designers can quickly fix their designs with less effort and mental demand compared to working through a large to-do list if people only accept feedback at the end. 
% Therefore, inspired by their findings, we aimed to help designers gather more quick feedback throughout the design process, keeping them engaged without overwhelming them.% and users' overreliance.

Despite these benefits, gathering constant feedback is often troublesome due to constraints like \textit{availability} and \textit{individuality} of the feedback providers~\cite{socialdynamics, cheng2020critique}. %When asking another person for design feedback, this is typically done in person or through online platforms. 
For example, one common way to gather quick design feedback is from personal network, a friend or close co-worker~\cite{kulkarni2015peerstudio}. They often have a strong connection with the designer and a good understanding of their background. %, and both likely have a good understanding of each other's background. 
However, there are social dynamics~\cite{socialdynamics} at play: feedback providers are not always \textit{available}, either physically (e.g., they are not in the same location or time zone) or socially (e.g., the designer feels bad to always ask the same person for feedback). Additionally, the pool of people able to provide frequent check-ins may not be diverse enough, in terms of background or \textit{individuality} (e.g., the feedback providers might have a limited variety of aesthetic preferences or lived experiences to support the designer effectively). Due to the close connection, feedback providers might also not give honest feedback to avoid damaging the relationship. %, while some designers may prefer receiving direct (and perhaps even `sassy') feedback at certain design stages. 
% Additionally, if they
% provide too much feedback, designers may become annoyed and anxious, thus disrupting their focus on the work. %For example, some designers seek constructive criticism to enhance the aesthetic or functionality of their design, while others want market or audience-specific insights to align their work with target demographics.

%Another way to seek quick feedback is through online forums or social media platforms~\cite{cheng2020critique, foong2017online}. According to our exploration on posts seeking feedback across such platforms, only those with substantial context—including both text and visuals—are likely to attract responses, as they provide sufficient background information and clarify designers’ goals. However, creating such context-rich posts  requires users to switch platforms, capture screenshots of their designs, and carefully compose both textual and visual explanations like design journey, problems,  and goals—adding to the time, effort, and back-and-forth communication. %In contrast, shorter posts tend to have a lower response rate~\cite{10.1145/2660398.2660427}.
%Thus, while feedback on these platforms can sometimes be high-quality and constructive, posting often requires significant effort, disrupts the design workflow, and responses may take a long time to arrive~\cite{krishna2021ready, foong2017online}. Therefore, we aim to make the feedback-gathering experience effortless and ambient, ensuring minimal disruption to the creative process. %Additionally, some platforms may not be easily accessible or could be behind paywalls. 

In this paper, we explore the usefulness of having an always-on ambient AI design companion that sits in the background, monitors the designer's workspace, and provides frequent, short-form, real-time feedback through various AI personas. Our aim was to investigate whether such an AI feedback companion could prove useful to designers, despite not having much intricate knowledge of what the designer is working on or aiming to achieve. We hypothesized that frequent real-time, read-aloud AI feedback that is always available---while perhaps not replacing human feedback---may still provide benefits to designers and help to alleviate the aforementioned issues when frequent high-quality human feedback is unavailable or hard to obtain.

Our AI design companion \systemname (Figure~\ref{fig:teaser}) uses the large language model (LLM) gpt-4-vision-preview~\cite{gpt4}, to analyze screenshots of the design workspace and generate tailored textual feedback. The screenshots, previous given feedback, and our template persona descriptions are combined into a feedback-generation prompt. This textual feedback is then sent to ElevenLabs' text-to-speech model~\cite{AIVoiceGenerator}, which generates a specific voice-over to read it out to users. This process is interactive, effortless, playful, accompanied, and iterative---drawing inspiration from the ``rubber duck debugging'' method used by designers and developers for problem-solving~\cite{thomas2019pragmatic}. With \systemname, users have a virtual ``rubber duck'' companion that offers frequent, real-time, read-aloud feedback. 

To explore how designers responded to quick AI feedback from \systemname, we conducted a design probe study with eight 3D Computer-Aided Design (CAD) designers and asked them to use the tool on one of their ongoing design projects. We decided to focus on 3D CAD design because it is a specialized yet rapidly evolving field with a wide variety of project types, allow us to understand how designers would use it to meet their unique workflows and design needs. Overall, participants shared positive experiences with \systemname, rating it as worth the effort, and appreciating how it provided them with a form of validation. A notable highlight was that participants felt less pressure when receiving AI feedback when compared to receiving human feedback. Despite some expected negative aspects of AI-generated feedback such as the AI's lack of context at times or issues with user control, our design probe study suggests that continuous AI-provided feedback has merits and could complement higher-quality human feedback, especially when human feedback is unavailable. Future improvements in AI models may also help to further narrow this gap.

In summary, this paper makes the following contributions:

% \begin{quote}
\begin{itemize}
    \item \systemname, a lightweight AI design companion that offers always-available, context-aware, ambient, and playful feedback experiences 
    through various personas, informed by an exploration of design feedback on 3D design forums. 
    \vspace{4px}
    \item Empirical results from a design probe study with eight 3D designers, demonstrating that \systemname is useful in providing inspiration and validation through ambient, convenient, and playful interactions. 
    \vspace{4px}
    \item{A discussion on the role of AI feedback, its strengths and weaknesses, and how to incorporate it into existing design workflows while balancing user involvement. We suggest that ambient interaction can be an valuable consideration for designing and evaluating future creativity support tools.}
\end{itemize}
% \end{quote}

\section{Related Works}

\subsection{Creativity, Review, and Feedback}

% Creative tasks are challenging. 
According to Hayes and Flower's cognitive process taxonomy over creative tasks, there are 3 stages: \textit{planning} (brainstorming, ideating), \textit{translating} (drafting, prototyping), and \textit{reviewing} ~\cite{FlowerHayes1981,hayesnew1996}. All three stages entail complex mental processes and impose demands on both the creative workers' short- and long-term memory, resulting in substantial cognitive load~\cite{hayesnew1996,Sparks}. Recent works have developed AI tools to support all three stages, especially the \textit{planning}~\cite{Petridis2023-oh,bursztyn_learning_2022,  kim_metaphorian_2023, 3dalle, disco, CreativeCoding} and \textit{translating}~\cite{visar,li_teach_2023,schick_peer_2022,yang_re3_2022,10.1145/3635636.3656189,petridis_constitutionmaker_2023,mirowski2022cowriting} stages.

The \textit{reviewing} stage includes evaluation (revisiting the creative artifacts) and revision (making iterations based on evaluations). 
Feedback plays an important role in evaluation. Thoughtful and diverse feedback can challenge preconceived notions, while providing validation and support~\cite{dannels2008beyond}. Feedback allows users to self-reflect and internalize external viewpoints, uncovering blind spots and leading to more innovative solutions alongside the original  goals~\cite{FlowerHayes1981}. Fundamentally, positive feedback throughout the design process may offer validations and support for designers, while constructive critical feedback from diverse providers helps refine the design, ultimately making it more understandable and appealing to a broader audience~\cite{socialdynamics}. Feedback does not need to be long to be effective --- as E et al.~\cite{janeEpaper}  mentioned, more continuous, short feedback gathered during the design process ensure designers make bigger improvements and encounter few issues in the final design compared to those who only received feedback at the end. Similarly, other studies~\cite{ullmanImportanceDrawingMechanical1990, kulkarni2015peerstudio} highlight designers' need for quick and continuous feedback throughout the design process. 
However, most AI tools for \textit{reviewing} focus on examining user interface requirements~\cite{10.1145/3654777.3676381, 10.1145/3613904.3642782, 10.1145/3613904.3642168}, supporting prompt engineering~\cite{chainforge, JohnnyPrompt}, paper writing and peer reviewing~\cite{10.1145/3640543.3645159,10.1145/3637371,10.1145/3613904.3642406,neshaeiEnhancingPeerReview2024,10.1145/3635636.3656201}, and practicing languages~\cite{sessler2023peer, 10.1145/3586182.3625114}. 
In comparison, relatively few works have focused on evaluating visual designs, including tasks involving more open-ended artistic vision or more mechanical requirements. Therefore, there is a gap in developing AI support tools for constant feedback gathering that can assist designers across all design phases.

\vspace{-10px}

\subsection{Ambient Technology}

Mark Weiser’s vision of ubiquitous computing~\cite{weiser_computer_nodate, weiserschool} imagined a future where computers are seamlessly embedded into all aspects of our homes and lives without being intrusive. The concepts of ambient technology~\cite{cook2009ambient} and calm technology~\cite{case_calm_2015} build on this idea, emphasizing that technology should promote calmness and avoid causing information or sensory overload. They offer benefits including  minimizing attention and distractions, enhancing productivity, and improving user satisfaction and immersion during use~\cite{kwok2024unobtrusive,10.1145/989863.989897}. An example is smart speakers like Amazon Alexa, which enable voice interaction to assist with daily tasks~\cite{long_challenges_2023}. These devices %don't take up much physical space, but 
enable hands-free and eyes-free interaction via voice-first commands, supporting the user in multitasking.  
As another example, Grammarly is a digital tool that accompanies users across platforms giving grammar feedback, fostering a sense of companionship~\cite{baileyWhatAmbientAutomation}. The design eliminates the need to switch platforms or copy and paste text for grammar feedback and editing. For creative tasks like writing, which involve high mental demand and already feature numerous visual cues on the screen, an ambient companion tool can enhance users' productivity and overall experience.

Most current creativity support research does not prioritize these ambient technology principles. Instead, there is a focus on standalone interfaces or distinctly separate steps within a design workflow~\cite{10.1145/3290605.3300619}. Separating the support functionality requires users to pause their current work and engage with the tool for assistance. This design is popular for emphasizing the `explorative' and `effort-reward trade-off' aspects, all prioritized by the Creativity Support Index~\cite{csi}. For example, a productive session with such systems often lasts at least 30 minutes before users can proceed~\cite{Tweetorial_ICCC, long_not_2024, ReelFramer}, and the tools often display an overwhelming amount of information dominating users' screen, designed to provide as many brainstorming or drafting supports as possible~\cite{10.1145/3491101.3503549,10.1145/3313831.3376739,10.1145/3582269.3615596}. They also provide clickable and interactive options to explore further, necessitating substantial visual focus and time commitment. In summary, this interface design is not entirely ideal for an ambient experience, as it requires a considerable portion of attention and engagement from users.
%On the other hand, ambient technology can lead to discoverability issues. Users may find it challenging to discern the full functionalities of a tool due to the lack of visual cues. Specifically, in creativity support tools, overly ambient and minimalist designs can cause confusion and a lack of control.
This highlights a need for developing ambient creativity support tools that seamlessly integrate into the design workflow without overwhelming designers~\cite{goncalvesEmergingOpportunitiesAmbient2017}.
\section{Exploration of Feedback on 3D Design Forums}

\begin{figure*}[b]
% \vspace{-10px}
    \centering
    \includegraphics[width=.68\linewidth]{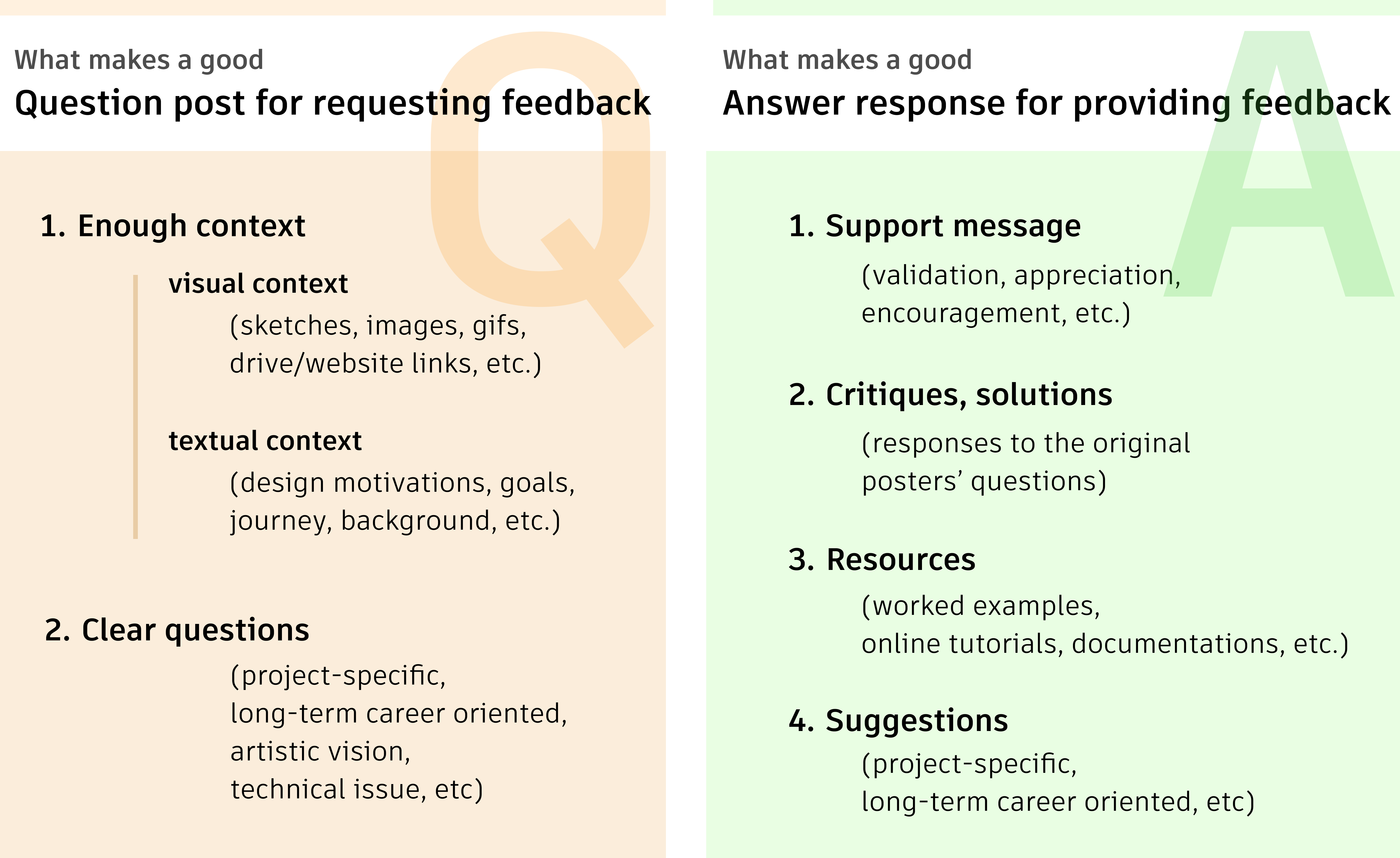}
    \caption{Framework based on the analysis of design forums, illustrating the components involved in \protect\tikz[baseline=-0.5ex] \protect\draw[fill=dot3, draw=black, line width=0.0mm] (0,0) circle (0.1cm); \textbf{(Q)}uestion posts for feedback requests (left, orange) and \protect\tikz[baseline=-0.5ex] \protect\draw[fill=dot1, draw=black, line width=0.0mm] (0,0) circle (0.1cm); \textbf{(A)}nswer posts for feedback responses (right, yellow). Question posts typically include visual context (sketches, images, and links), textual context (design motivations, goals, and background), and various types of questions (ranging from project-specific to career-oriented). Answer posts, responses, usually offer supportive messages (validation and encouragement), critiques and solutions addressing the posted questions, resources (worked examples and tutorials), and suggestions tailored to the project or career goals.}
\label{fig:feedbackmodelelicitedfromformative}
\Description[A two-column table presenting our elicited framework from the analysis of design forums.]{A two-column table presenting our elicited framework from the analysis of design forums. The two columns illustrate the components involved in (Q)uestion posts for feedback requests (left, orange) and (A)nswer posts for feedback responses (right, yellow). The caption includes all the necessary information.}
\end{figure*}

Nowadays, people seek design feedback from online forums in addition to their personal networks, but such resources also have important limitations~\cite{socialdynamics, cheng2020critique}. Thus, we conducted a formative exploration on various feedback platforms for 3D art and 3D CAD designers, including Reddit, Discord, and designer forums, to abstract the current human-to-human feedback-gathering process and identify design guidelines for our tool.

We focused on Polycount.com\footnote{\href{https://polycount.com/forum}{https://polycount.com/forum}}, one of the largest 3D art and CAD design forums that serves both professional and hobbyist artists specializing in creating 3D art for video games. The platform is often chosen for its active community and its ability to reflect artists’ perspectives and dynamics in depth~\cite{chenMemoVisGenAIPoweredTool2024,gallagherEmbedded3DWeb}. 
After filtering and analyzing Polycount posts over several consecutive days in early June 2024, we identified and selected ten feedback-seeking posts that appeared at the top of the time-ranked feed on June 18, 2024, based on specific criteria:  having five or more new replies (excluding replies from the original posters), and including explicit questions aimed at soliciting feedback rather than showcasing.
Our analysis, using grounded theory~\cite{strauss1994grounded} to identify patterns in the feedback-seeking process, results in the framework presented in Figure  \ref{fig:feedbackmodelelicitedfromformative}. This framework outlines the typical elements found in the posts, specifically the questions (Q) asked and the answers (A) provided. %We examined the types of questions designers ask, such as critiques with project-specific engineering details or about high-level artistic vision, and requests for technical issues solutions or career-oriented suggestions. We also analyzed how these questions are formulated, including the accompanying design files and contextual information. Furthermore, we explored the forms of feedback given, which include support messages, critique, solutions, explanations, external resources, and project-specific or high-level career advice. 
Then, we verified this framework by applying it to posts on popular 3D-related subreddits on Reddit, including r/design\_critique (105K members)\footnote{\href{https://www.reddit.com/r/design_critiques/}{https://www.reddit.com/r/design\_critiques/}}, r/3Dmodeling (1.3M members)\footnote{\href{https://www.reddit.com/r/3Dmodeling/}{https://www.reddit.com/r/3dmodeling/}}, and r/ArtCrit (105K members)\footnote{\href{https://www.reddit.com/r/ArtCrit/}{https://www.reddit.com/r/artcrit/}},  as well as the ``3D Modeling'' Discord server\footnote{\href{https://discord.com/invite/Dayy7kFgJG}{https://discord.com/invite/Dayy7kFgJG}}, which had approximately 26K members at the time of June 19, 2024. We examined over 60 top Reddit posts and past Discord channel conversations across platforms from June 19 to June 24, 2024. Our findings indicate that the framework of asking questions and providing feedback is similar across platforms and time. Feedback on Discord and Reddit tends to be less formal, shorter, and has an slower turnover rate. This analysis provides a comprehensive framework of the feedback exchange process, highlighting gaps, such as the extensive components required in a post, long response latency, the need to cross-post or re-post to ensure sufficient feedback. Our analysis identifies three challenges and we propose six corresponding design guidelines (DGs).

\vspace{7px}
\textbf{Challenge 1: Slow responses}; and \textbf{Challenge 2: High effort required.} %Feedback-seeking posts receive responses slowly while original posters spend great efforts creating them.
We observed that most posts on Reddit, Discord, and Polycount remain unresponded to within 24 hours, and when they do receive responses, the feedback is sometimes unhelpful. Additionally, original posters put significant effort into including extensive contextual information in their posts: screenshots of their design or a link to their near-finished portfolio, a few paragraphs about their design journey (including design goals, context, progress), and a few lines about their requests (including seeking low-level technical solutions, high-level artistic vision critiques, external tutorials or resources, or even just validation and support). For example, nine out of the ten Polycount posts we analyzed exceeded 90 words per question post, with an average length of 165.4 words. Three posts included links to personal drives or portfolios, while the other seven attached at least two media files (average five files), such as screenshots, GIFs of their designs, or reference photos.  %\\tao{DG A SHORT CALCULATION IF HAVE TIME :Besides, on average, feedback-seeking posts recieve their first reply from others after X hours on Polycount, X hours on Reddit, X hours on Discord}  tao{should i also mention its hard to find human next door / irl to give feedback? or only mentioning online is fine...} 
These findings align with prior work that highlights the barriers to receiving creative feedback online ~\cite{socialdynamics, cheng2020critique, kulkarni2015peerstudio, portfoliomentor}.  \textbf{Opportunity}:  There is room for a tool to provide instant on-demand feedback. The tool can function as a consistently accessible companion for users throughout the design workflow. It should optimize the feedback-gathering process to enhance time efficiency and minimize disruption to the designer's workflow. This includes providing quick feedback access without requiring users to navigate to a website, input text questions, or manually take screenshots. Additionally, the tool should foster an encouraging and effortless experience by reducing both the physical and cognitive effort involved in crafting question posts and alleviating concerns about social judgment from the platform audiences. %This approach makes the feedback process more accessible.
Consequently, we identified four design guidelines for our tool's \textit{feedback interaction}: 

\begin{itemize}
% \vspace{-4px}
    \vspace{3px}
    \item \textbf{DG1) Always available}: The tool should be omnipresent, offering users reliable and immediate access to high-quality feedback throughout all stages of design.
    \vspace{3px}
    \item \textbf{DG2) Context-aware}: It should automatically incorporate necessary design context materials for generating feedback without requiring users to manually input. The tool should include both visual (screenshots) and textual context (previously given feedback) while generating new feedback.
     \vspace{3px}
    \item \textbf{DG3) Ambient}: The tool should be seamlessly integrated into the design workflow, allowing users to access feedback without changing applications or taking additional steps.  It should facilitate a calm experience by delivering feedback in a way that avoids disturbance or information overload.
     \vspace{3px}
    \item \textbf{DG4) Playful}: The tool should foster a positive experience by making the feedback process more interactive and enjoyable, reducing anxiety associated with feedback-gathering.
\end{itemize}

\vspace{2px}

\textbf{Challenge 3: Missing multiple perspectives.} Many feedback-seeking posts only receive a very limited number of responses. % aim to obtain critical critiques or actionable suggestions to improve the project. However, as noted earlier, these posts often receive only a limited number of responses. 
Some may focus on other designers' shared experiences or questions without providing constructive feedback, while some includes sarcasm or discouragement, resulting in few helpful replies. Even when feedback is of high quality, the fundamental scarcity of design experts on such platforms often leads to a lack of diversity and varied perspectives. Thus, designers find it necessary to repeatedly post their requests across multiple platforms and at different times to gather sufficient and distinct perspectives. 
This aligns with works on how and why creative professionals need diverse perspectives in \cite{bhattacharjeeUnderstandingCommunicationPreferences2024,hui2014crowd,yen2016social}.
%\cite{10.1145/3613904.3642406} did persona work. 
% \tao{Recent works, including ChoiceMates~\cite{parkChoiceMatesSupportingUnfamiliar2025}, Impressona~\cite{10.1145/3613904.3642406}, and Proxona~\cite{choiProxonaSupportingCreators2025} design conversational agent personas with different types of \textit{expertise} to uniquely support users in decision-making. Khadpe et al.’s conceptual metaphors and Bhattacharjee’s work~\cite{10.1145/3415234, bhattacharjeeUnderstandingCommunicationPreferences2024} on dimensions of \textit{warmth} and \textit{communication traits} show how varying interpersonal styles and linguistic tones can significantly influence how feedback is perceived and received.}
\textbf{Opportunity:} There is an opportunity for a tool capable of providing feedback from diverse perspectives specifically for design, offering an interactive experience that allows users to receive varying feedback as needed. We identify two design guidelines for our tool's  \textit{feedback content}:

\begin{figure*}[b]
    \centering
    % \vspace{-10px}
    \includegraphics[width=\linewidth]{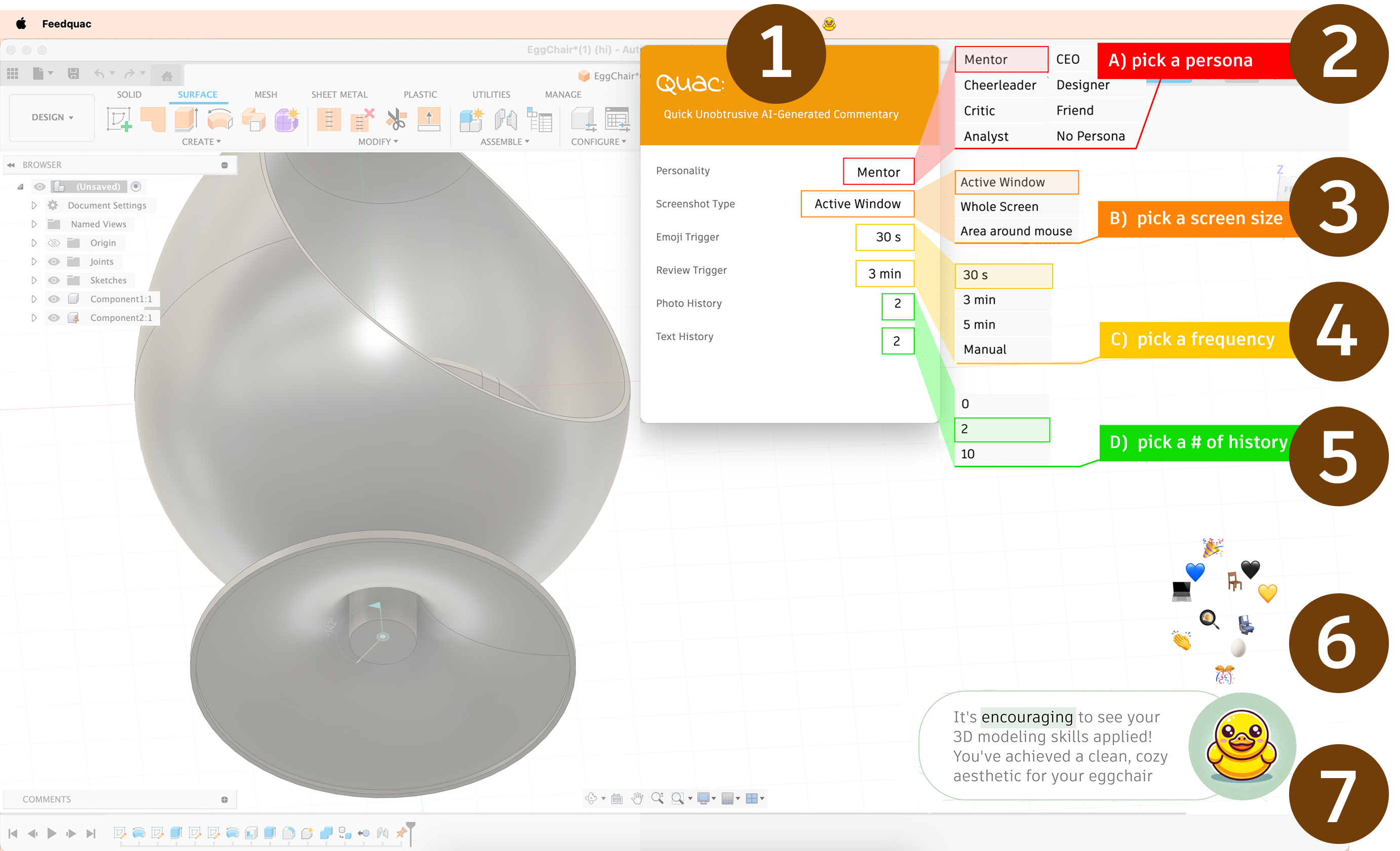}
    \caption{Annotation \protect\systemname User Interface Walkthrough: Tanya, a 23-year-old entry-level 3D CAD designer, is designing her first egg chair in Fusion 360. Unsure about the curvature of the chair's bottom base, she missed her TA’s office hours and finds online feedback too slow. She turns to \protect\systemname for real-time guidance. To start, she opens \protect\systemname and accesses the control panel \protect\filledcircnum{1} for setup. She selects the TA-like constructive \textit{Mentor} persona for feedback \protect\filledcircnum{2} and chooses to capture the entire active window of Fusion 360 \protect\filledcircnum{3}, avoiding distractions from other open applications. For frequency, she sets emoji feedback every 30 seconds and audio feedback every 3 minutes \protect\filledcircnum{4}. She also ensures the system retains the last two feedback generation as part of the context to avoid repetition \protect\filledcircnum{5}. As she works, a set of egg and chair emojis appears from the duck icon, humorously acknowledging her design \protect\filledcircnum{6}. Curious, she manually triggers feedback using \texttt{Command + R}. The duck icon at the bottom right rotates, and after 5 seconds, a supportive female voice encourages her progress: ``\textit{It’s encouraging to see your 3D modeling skills applied... Considering refining the base...}'' \protect\filledcircnum{7} The spoken feedback also appears as text next to the icon. Encouraged, Tanya adjusts the chair’s bottom base shape. Three minutes later, without manual pressing the hotkeys, the duck automatically rotates again, delivering another round of mentor-like supportive feedback. }
    \label{fig:systemUI}
\Description[An annotated screenshot of the FeedQUAC user interface]{
An annotated screenshot of the FeedQUAC user interface. The control panel is labeled 1. The selection of the Supportive TA-like Mentor persona is marked 2. The option to capture the entire active window of Fusion 360 is indicated by 3. The frequency settings for emoji feedback every 30 seconds and audio feedback every 3 minutes are shown at 4. The setting to retain the last two feedback generations for context is labeled 5. Egg and chair emojis appearing from the duck icon are annotated at 6. A manually triggered feedback action using Command + R is noted at 7, where the duck icon rotates, and spoken feedback is displayed as text.}
\end{figure*}

\begin{itemize}
 \vspace{3px}
    \item \textbf{DG5) Diverse feedback}: The tool should deliver feedback encompassing various perspectives, personality, and expertise, to provide comprehensive insights for users.
     \vspace{3px}
    \item \textbf{DG6) Iteration-aware  feedback}: The tool should provide feedback that adapts to the iterative nature of the design process. Users can request feedback on specific areas of their design rather than the entire project. The tool should also be able to identify and provide insights on recent adjustments made by users throughout the design process.
\end{itemize}

\newpage
% \newpage

\newpage

\section{System}

\subsection{System Design}

According to the exploration of feedback on design forums, we designed an interactive app that floats at the top of the designers' editor screen to provide feedback for users to interact with. The tool follows the design guidelines (DG1-7) as always available, context-aware, ambient, playful, and contains a variety of responses. Following the vision of rubber duck debugging~\cite{thomas2019pragmatic}, we want our tool to help people find issues in their design and gain inspiration. Thus, we designed this tool along with the visuals of a duck.

\begin{figure*}[!b]
    \centering
    % \vspace{-10px}
    \includegraphics[width=\linewidth]{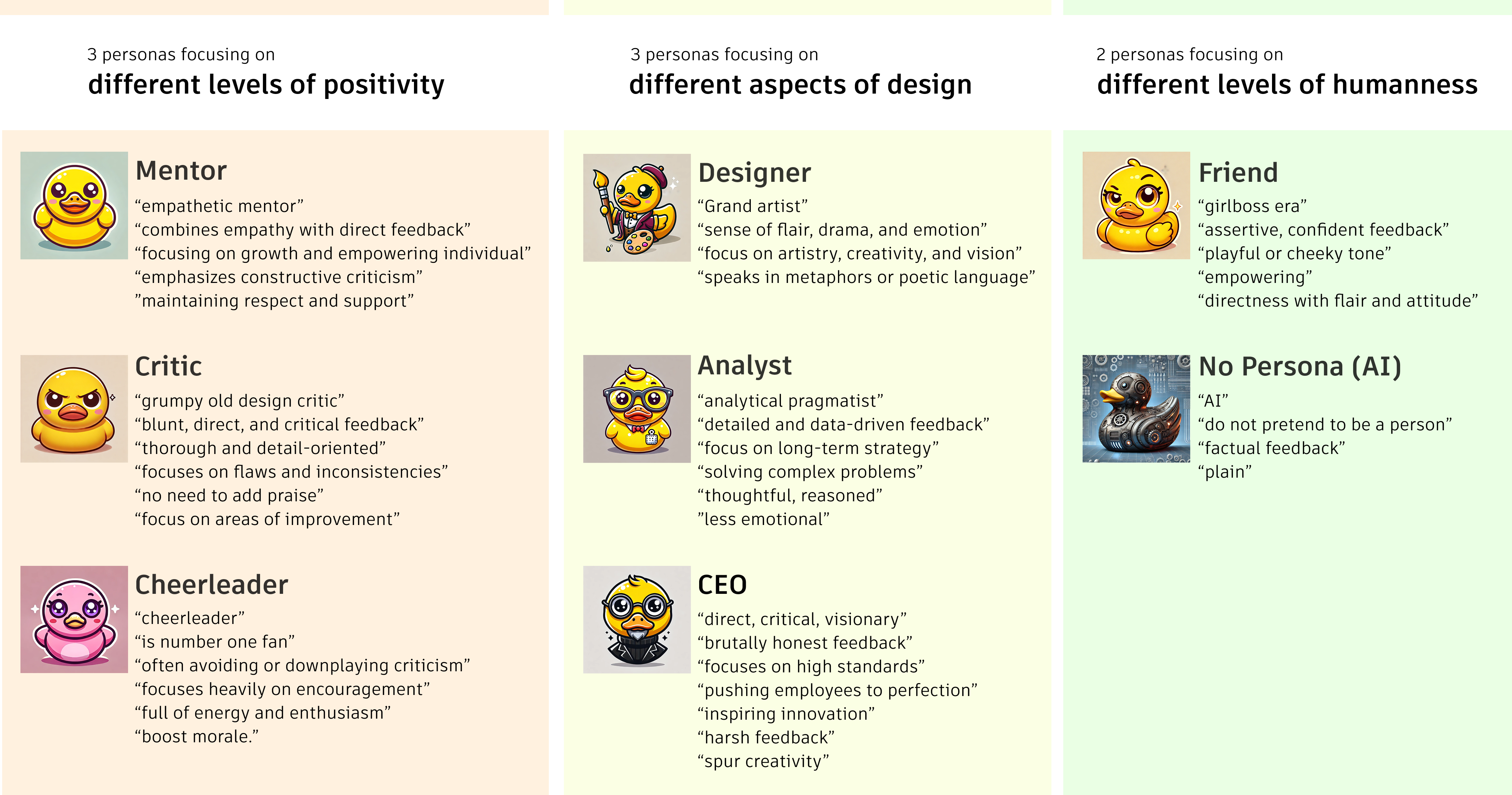}
    \caption{An illustration of the eight personas used in \protect\systemname, along with their icons and corresponding keywords used in the personality prompts. The three personas on the leftmost orange column cover {\protect\tikz[baseline=-0.5ex] \protect\draw[fill=dot3, draw=black, line width=0.0mm] (0,0) circle (0.1cm); different levels of positivity}: constructive \textit{Mentor}, supportive \textit{Cheerleader}, and grumpy \textit{Critic}. The three personas in the middle yellow column focus on {\protect\tikz[baseline=-0.5ex] \protect\draw[fill=dot2, draw=black, line width=0.0mm] (0,0) circle (0.1cm); different aspects of design}: aesthetic \textit{Designer}, practical \textit{Analyst}, and innovative \textit{CEO}. The two on the rightmost green column represent {\protect\tikz[baseline=-0.5ex] \protect\draw[fill=dot1, draw=black, line width=0.0mm] (0,0)circle (0.1cm); different humanness levels}: sassy \textit{Friend} and robot-like \textit{No Persona (AI)}. See the full gpt-4-vision-preview prompts and ElevenLabs voice IDs and descriptions in Appendix A.1 and A.2. See examples of each persona's feedback in Appendix A.3.}
    \label{fig:PersonaPreview}
    \Description[An illustration of the eight personas used in FeedQUAC]{An illustration of the eight personas used in FeedQUAC, along with their icons and corresponding prompt keywords used in the personality prompts. The caption and appendix A.1 includes all the necessary information.}
    % \vspace{-10px}
\end{figure*}

% We designed this tool to be always available and ambient to use. 
To ensure the \textbf{always available (DG1)}  interaction, we made the tool represented by a floating duck icon at the bottom right of the users' screen, as shown in Figure  \ref{fig:systemUI}. The duck icon always float over the user's design editor and screen, making it easy to locate. We designed the whole feedback gathering experience to be as seamless as possible to satisfy the \textbf{ambient (DG3)} design goal. Users can interact with the tool using keyboard shortcuts and receive feedback both read aloud and displayed beside the duck icon. We provide two types of feedback to support the design process. The first is a lightweight emoji-based validation, designed for an extreme-ambient, text-free experience. This feedback provides supportive and contextually relevant emojis to acknowledge the user's design progress, activated with \texttt{Command + E} on Mac or \texttt{Windows + E} on Windows. The second is a full-length voice review, which delivers a short yet detailed critique of the design or recent modifications. Users can trigger this feedback manually using \texttt{Command + R} on Mac or \texttt{Windows + R} on Windows. As a voice feedback is requested and then read aloud, a voice transcript with highlights appears next to the floating duck icon. %, allowing users to choose their preferred feedback consumption method. 
By integrating shortcut keys and voice-based interactions, users can avoid switching from their design editor to a standalone website, thus maintaining workflow continuity without being overwhelmed by textual information. 
%\tao{omg the whole emoji stuffs... tbh the interview and survey have really little results for emojis, should we just not mention it?} 
If users wish to request different feedback, they can press the shortcut key at any time. The tool also allows users to set an automatic feedback-seeking cycle (every 30 seconds, 3 minutes, or 5 minutes) and offers the option to turn it off. Such a cycle for automatic feedback generation allow users to concentrate on design while receiving feedback at predetermined intervals without pressing hotkeys, thereby minimizing disturbances and preventing cognitive overload.

The tool is \textbf{context-aware (DG2)}, ensuring users incorporate necessary contexts for feedback generation effortlessly. In our design, context-gathering occurs automatically after users pressed the shortcut key: the tool captures a screenshot of the current design on screen (visual context) and send to AI models to generates feedback. Users can select from three screenshot types: whole screen, active window, or the area around the cursor. Additionally, the system can incorporate previously taken screenshots and received feedback (textual context)  as part of its memory, acknowledging users' design progress and avoiding repetition. Users can adjust the memory level to their preference. In generating \textbf{iteration-aware feedback (DG6)}, users can configure feedback settings---such as feedback cycle frequency, screenshot type, and memory of previous feedback and screenshots---by clicking the small duck icon at the top of the screen before generating more feedback.

% \newpage
The tool incorporates various personas to generate inspirational and \textbf{diverse feedback (DG5)}. We introduced eight initial feedback personas to encourage a variety of feedback styles, as listed in Figure~\ref{fig:PersonaPreview}. These personas were derived from our formative assessments and an initial brainstorming session: the authors co-identified several axes for variation, including supportiveness, focus, and humanness. We elicit these axes based on the prior works in conversational agent design, which provided personality-driven and linguistic cue-driven feedback~\cite{parkChoiceMatesSupportingUnfamiliar2025,10.1145/3613904.3642406,choiProxonaSupportingCreators2025, 10.1145/3415234, bhattacharjeeUnderstandingCommunicationPreferences2024}. While we discussed allowing the user to configure their own persona by selecting different options along each axis, we opted for a pre-configured list of eight for simplicity and comparison’s sake. When choosing our personas we tried to select a balance from across the design space. 
% We selected three personas focused on different positivity levels (constructive \textit{Mentor}, supportive \textit{Cheerleader}, grumpy \textit{Critic}), three that focused on different aspects of the design (practical \textit{Analyst}, innovative \textit{CEO}, aesthetic \textit{Designer}), and two levels of humanness (sassy \textit{Friend}, robot-like \textit{No Persona (AI))}. 
First, we included three personas focused on {\protect\tikz[baseline=-0.5ex] \protect\draw[fill=dot3, draw=black, line width=0.0mm] (0,0) circle (0.1cm);} different positivity levels (constructive \textit{Mentor}, supportive \textit{Cheerleader}, grumpy \textit{Critic}). This dimension was inspired by Khadpe et al.’s conceptual metaphors and Bhattacharjee’s work~\cite{10.1145/3415234, bhattacharjeeUnderstandingCommunicationPreferences2024} on dimensions of \textit{warmth} and \textit{communication traits}. They both illustrate how personality-driven feedback—featuring varying interpersonal styles, warmth, and linguistic tone—can significantly influence how feedback is perceived and received.%Bhattacharjee et al.’s work~\cite{bhattacharjeeUnderstandingCommunicationPreferences2024}, which investigates user preferences for feedback tone and warmth and found that personality-driven language in conversational agents affects user engagement and receptivity.
Second, we included three personas that focused on {\protect\tikz[baseline=-0.5ex] \protect\draw[fill=dot2, draw=black, line width=0.0mm] (0,0) circle (0.1cm);} different aspects of the design (practical \textit{Analyst}, innovative \textit{CEO}, aesthetic \textit{Designer}), building on prior work like 
%ChoiceMates~\cite{parkChoiceMatesSupportingUnfamiliar2025} and Proxona~\cite{choiProxonaSupportingCreators2025}, which emphasize that agents with tailored decision roles or creative functions can better support user needs during ideation and evaluation.
ChoiceMates~\cite{parkChoiceMatesSupportingUnfamiliar2025}, Impressona~\cite{10.1145/3613904.3642406}, and Proxona~\cite{choiProxonaSupportingCreators2025}, which emphasize that agents with tailored decision roles or creative expertise can better support user needs during ideation and evaluation.
Third, we included two on the rightmost column represent {\protect\tikz[baseline=-0.5ex] \protect\draw[fill=dot1, draw=black, line width=0.0mm] (0,0) circle (0.1cm);} different humanness levels (sassy \textit{Friend}, robot-like \textit{No Persona (AI)}). This axis draws on Khadpe et al.’s metaphor-based framing~\cite{10.1145/3415234}, which explores how warmth and competence shape perception of agents, and Doyle et al.~\cite{doyleMappingPerceptionsHumanness2019}, who demonstrate how anthropomorphic traits influence user expectations and interaction styles.

Each persona is guided by a personality prompt (with keywords listed in Figure  \ref{fig:PersonaPreview} and full details in Appendix A), followed by a general feedback generation prompt introducing the feedback task and attaching screenshots, and finally an  context prompt, which incorporates previously given feedback to avoid repetition. Every time when users request feedback, either within the automatic feedback cycle or by manually pressing the shortcut keys, the three prompts and encoded screenshots are sent to a language model.
Also, to enhance session \textbf{playfulness (DG4)} and distinguish between personas, we created diverse duck avatars representing different personas, generated using ChatGPT. We first generated the main icon and then reprompted ChatGPT for variations based on the other persona keywords. This variety underscores the playful nature of the interactive characters. The voices chosen for the personas also exhibit significant diversity, including a ``middle-aged English non-binary'' with a grumpy tone\footnote{ElevenLabs Voice ID: \texttt{O7p2vmz2iEYgMXxkbsif}. \href{https://elevenlabs.io/app/voice-lab?voiceId=O7p2vmz2iEYgMXxkbsif}{Preview link: \faIcon{link}}}) for the grumpy \textit{Critic} persona, a ``sassy young female'' with a sarcastic tone\footnote{ElevenLabs Voice ID: \texttt{jsCqWAovK2LkecY7zXl4}. \href{https://elevenlabs.io/app/voice-lab?voiceId=jsCqWAovK2LkecY7zXl4}{Preview link: \faIcon{link}}}) for the sassy \textit{Friend} persona, etc. After testing, we refined the feedback prompts and voice selections for all eight personas (See full gpt-4-vision-preview prompts, ElevenLabs voiceIDs and descriptions in Appendix A).

\subsection{System  Implementation}
% \tao{NEED A FINAL INTRODUCTION ON PROMPTS. EMOJI PROMPTS, FEEDBAC KPROMPT, MANUAL/AUTO}
% \tao{Section 3.3: This still sounds like there's very little system contribution, it's just plain use of gpt-4-vision-preview and ElevenLabs. If that's the case, that's fine, but we should then clarify what aspect of \systemname is a contribution, if any. Some of the terms are a little vague (simple and lightweight) and need more explanation (what is the persona prompt?).}
The system is implemented as a cross-platform desktop application using JavaScript and Electron. %To ensure that the tool won't make the design experience slower, we developed the tool to be simple and lightweight as a floating tool on users' editor screens. 
To realize the design guidelines, the tool is designed to be lightweight and minimal research prototype, functioning as a floating element on users' editor screens.
Upon a user's request for feedback from keyboard, the tool automatically captures a screenshot of the user's design, encodes it as a base64 image, and integrates it into the prompt sent to a language model. In this study, we utilized OpenAI's gpt-4-vision-preview API~\cite{OpenAIPlatform} (access date for user study: September 30 - October 2, 2024), along with the persona prompt, to generate textual feedback. We then used ElevenLab's text-to-speech model API~\cite{AIVoiceGenerator}, along with the persona's chosen ElevenLab voice ID, to produce voice feedback as an .mp3 file. The feedback is played to the user, with the entire process (from \texttt{COMMAND + R}, the start of the feedback cycle) taking approximately 5-10 seconds. All feedback generation activities are logged on the experimenter's laptop during experiments.
Our tool is agnostic to which AI we actually use, allowing for future fine-tuning and integration, like retrieval-augmented generation (RAG) or plugging in newer AI models to enhance its capabilities.

\section{Design Probe Study}
%\tao{Section 4.1: The mentioned areas of art, product design, game development etc. don't really relate to Fusion 360 it seems? }
%As \systemname offers vairous apply in their 3D CAD design projects, we conduct a design probe study with eight 3D designers. 
As \systemname offers various potential applications across different stages of 3D CAD design projects, we conducted a design probe study with eight 3D CAD designers. 
%We specifically chose to focus on 3D CAD design because it is a specialized yet rapidly evolving field with a wide variety of project types. %This diversity provides a unique opportunity to examine the potential applications and benefits of our tool. 
This approach allowed us to uncover the tool’s  diverse usage patterns and understand how designers can use it to meet their unique workflows, design needs, and project complexities. %By observing how 3D designers interact with our tool, we aim to gain valuable insights into how our tool's interaction and feedback content design can support and enhance their design process.

\subsection{Demographics} 
We recruited eight 3D designers using on-campus posters, 3D designer community Slack groups, and campus-wide mailing lists. %We initially received more than 100 sign-ups but screened out participants who didn't meet our criteria. 
All participants were deemed experienced 3D designers, as they have at least two to nine years of 3D design experience and have completed at least three projects using  \idename, the design editor we chose to run experiments on. %, in areas such as art, product design, game development, architectural visualization, or animation.
% Out of the eight participants, five are male, two are female, and one is identified as male, nonbinary, genderqueer, and genderfluid. With an average age of 21.13, they have an average of 4.63 years of experience in 3D design. Among them, three participants identified as senior-level, four as mid-level, and one as entry-level designer. 
All eight participants have a good understanding of generative AI, with six sharing that they have read about it and grasp the key concepts, while the other two reported having in-depth knowledge of generative AI models and tools.
Five participants use generative AI %\footnote{In the demographic survey, we specify the generative AI usage as using language models (ChatGPT, Gemini, Claude etc), text-to-image or image-to-text models (DALL-E, Midjourney, Adobe Firefly etc), or text-to-speech models.} 
weekly or monthly, one uses daily, and two uses it less than once a month. Seven participants have used generative AI in their design process, primarily for ideation or brainstorming (six participants), for enhancing or refining existing designs (three), and for automating repetitive tasks like resizing or color adjustments (three).
Finally, five participants hold positive attitudes towards integrating generative AI into their design process, while three shared a neutral stance.
 % We explored how they initially accept and understand the tool, how they integrate it into their own processes, and, based on their current methods of soliciting and utilizing (AI) feedback, what opportunities exist for our tool to support designers. Finally, we examined potential paradigms for new ambient creativity support tool interactions.

% \begin{figure}[!h]
%     \centering
%     \includegraphics[width=1\linewidth]{fig/demographicsTable.png}
%     \caption{Enter Caption}
%     \label{fig:enter-label}
% \end{figure}

\subsection{Procedure} 
%\tao{Section 4.2: I would add the actual questions in an appendix.}
The 70-minute design probe study took place over Zoom, with participants granted remote control to access \systemname on the experimenter’s laptop. Before the study, we asked the participants to share the \idename~.f3d files of either one of their ongoing projects they were comfortable working on and sharing with us.
During the study, participants were first introduced to the study and asked for consent. We then conducted a short interview about the participants’ past experiences with 3D design and feedback gathering. After that, we presented \systemname with a 5-minute demo video showing the basic functionality and use cases. We then asked the participants to continue their ongoing design project at least 25 minutes using our tool\footnote{P2 did not complete the full 25-minute design task due to internet connection issue, but their experience was well represented as we confirmed with them during interview.}. Throughout such design task, every time they received feedback from the tool, the experimenter would ask their thoughts on it and encourage a think-aloud protocol. Finally, we asked them to fill out a post-study survey and participate in an interview to gather demographic data and understand their detailed experience. We asked questions about their perceptions, expectations, quality, user control, and interaction with our system, and how they felt the interaction benefited their original design and feedback-gathering workflow. Each participant was offered a \$75 payment for their time.
The study protocols were evaluated and approved by our internal ethics review committee (see full protocols in Appendix B and C).

\begin{figure*}[!b]
    \centering
    \vspace{5px}
    \includegraphics[width=1.0\linewidth]{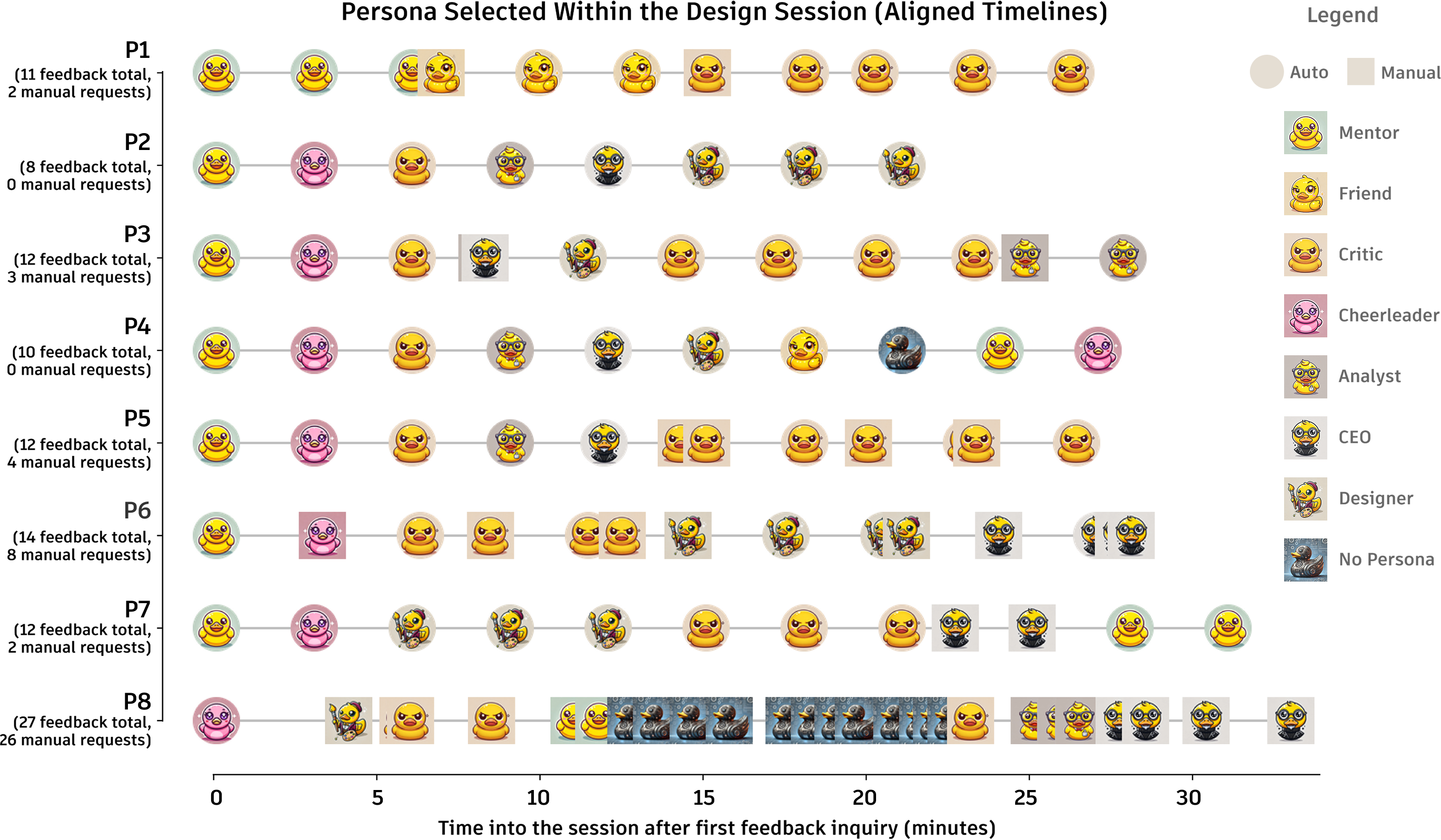}
    \caption{Visualization of feedback requests for each participant during the design sessions, showing the selected persona, the timing of feedback requests, and the mode of generation---either \protect\tikz\fill[black] (0,0) circle (0.08); automatic (within the three-minute feedback cycle, represented as circles) or \protect\(\blacksquare\) manual (via \texttt{Command + R}, represented as squares) . The timeline for each participant is aligned with their first feedback request for consistency in comparison. On average, participants received 13.25 feedback instances per session, with 5.26 instances manually requested and the remaining 7.99 instances generated automatically.  Some participants, such as P2 and P4, relied solely on the automatic feedback cycle, while others, like P6 and P8, requested manual feedback via hotkeys for more than half of their feedback requests. These participants also manually explored different personas after reviewing the early feedback.  Among all persona requests, \textit{Critic} (31 times) was the most frequently used persona.The most frequently manually requested personas were \textit{No Persona (AI)} (12 times) and \textit{Critic} (11 times).}
    \label{fig:logVisualizationDucks}
     \Description[An illustration of the participants’ feedback requests during design sessions]{An illustration of the participants’ feedback requests during design sessions, detailing the timestamps of feedback requests, the selected persona, and the feedback generation modes. P1 requested 11 total feedbacks, with 2 manual requests; P2 requested 8 total feedbacks, with 0 manual requests; P3 requested 11 total feedbacks, with 3 manual requests; P4 requested 12 total feedbacks, with 4 manual requests; P5 requested 12 total feedbacks, with 4 manual requests; P6 requested 14 total feedbacks, with 8 manual requests; P7 requested 12 total feedbacks, with 2 manual requests; P8 requested 27 total feedbacks, with 26 manual requests. Among the feedback personas, the following section 6.1.2 includes all the necessary information.}
\end{figure*}

\subsection{Data Analysis} 
%\tao{Section 4.3: I don't think we want to do t-tests and statistical analysis on Likert scale data. I would just visualize it and analyze it that way and strengthen it with qualitative data. }
We used Python to clean the logs and extract user interactions with our tool, including timestamps for feedback requests — either automatic or manual — and the associated feedback data (input screenshots, selected persona types, and feedback output). For the survey data, we created diverging stacked barcharts to visualize the responses distribution. For the interview data, we performed an inductive thematic analysis and created affinity diagrams using grounded theory~\cite{cooper_thematic_2012, strauss1994grounded} from the think-aloud and interview recording transcripts. 
The themes generated from the affinity diagrams were presented to all the authors for feedback.
% \newpage
% \vspace{10px}
\section{Results}

\subsection{Log Analysis}

\subsubsection{\textbf{Frequency}} Eight participants received a total of 106 pieces of feedback, with 45 of those being manually triggered and 61 triggered by a three-minute timer (See Figure  \ref{fig:logVisualizationDucks}). On average, a participant engaging in a 25-minute design session with \systemname received feedback 13.25 times. Of these, there were on average 5.26 instances that were manually requested via \texttt{Command + R}. 

We observed varying patterns of feedback requests among different participant groups during the study. First, participants like P1, P3, P5, P6, and P7 showed similar usage patterns where they requested manual feedback at least twice or more after reviewing some early feedback. On average, the first manual feedback request occurred around 11.02 minutes into the session (after the first feedback). This group of participants manually requested feedback an average of 3.6 times per session. In contrast, P2 and P4 never manually requested feedback throughout their sessions, instead focusing exclusively on their design projects. P8 did not turn on the automatic feedback generation, unlike the other seven users, who used the 3-minute automatic feedback cycle at least one time. Instead, they frequently press shortcut key \texttt{Command + R} for manual design feedback, nearly every minute after initial exposure, totaling 27 manual feedback requests during the study period.

\subsubsection{\textbf{Types of feedback persona}} All eight personas are used, while the \textit{Critic} persona is used most frequently -- 31 times. %(taking 29.2\% of the overall generation). 
The other popular personas are \textit{CEO} (14 times), \textit{Mentor} (14), and \textit{Designer} (13). As mentioned in the previous section, those participants who manually requested feedback needed to undergo a set of explorations in order to find their desired persona. For example, P1, P3, and P7 kept the automatic generation pattern but set their persona to design critic after a few rounds of trials. In comparison, P5 and P8 tried to request feedback from the \textit{Critic} for later manual feedback requests. P6, P7, and P8 specifically requested manual feedback from the \textit{CEO}  and \textit{No persona (AI)} for feedback. Out of the manual requests, we found that people specifically requested feedback from \textit{No persona (AI)} (12 times), \textit{Critic} (11), \textit{CEO} (10), and \textit{Analyst} (5).  %%address geroge's comment: 
The variation in persona usage suggests that the number of personas needed may depend on the individual, raising an open question about how many personas are optimal for design feedback and whether this number varies across users. 
\subsection{User Experience}

% \newline

Six out of the eight participants agreed that they had somewhat positive to positive experiences with \systemname (Figure  \ref{fig:overallexperience}a). All eight participants found receiving feedback from \systemname to be low-stakes and stress-free (Figure  \ref{fig:overallexperience}b). In contrast, only three participants found receiving human feedback low-stakes (Figure  \ref{fig:overallexperience}c). Seven participants perceived human feedback as more high-stakes than AI feedback. These findings suggest that \systemname can be a more approachable and low-stress alternative for receiving feedback. In addition, all eight participants reported having made progress on their design project during the study.

\begin{figure}[!b]
\vspace{-10px}
    \centering
    \includegraphics[width=1\linewidth]{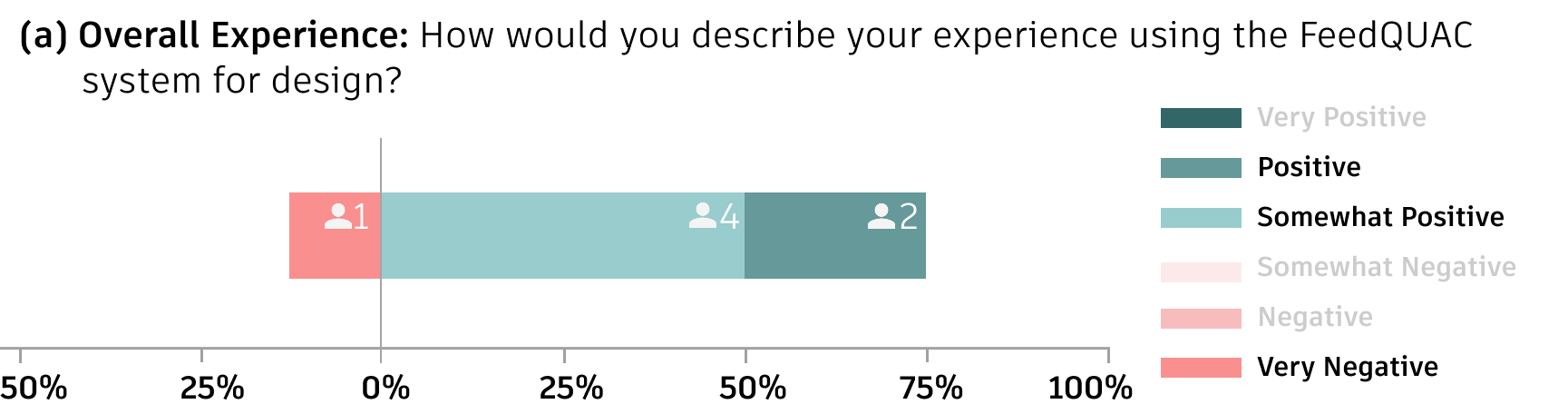}
    
    \vspace{7.5px}
    \includegraphics[width=1\linewidth]{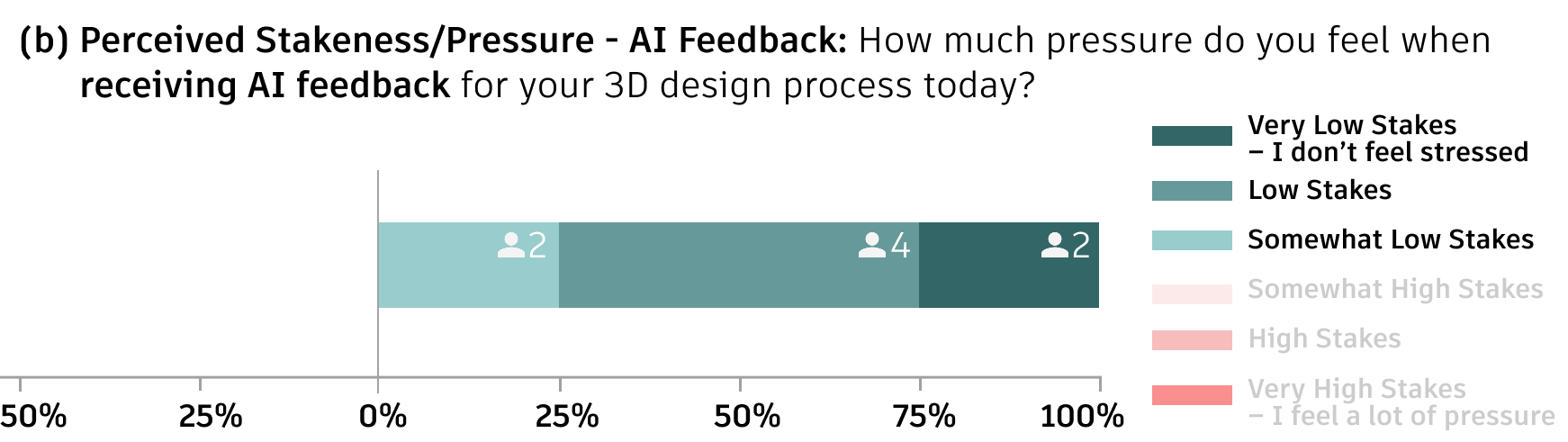}
    
    \vspace{7.5px}
    \includegraphics[width=1\linewidth]{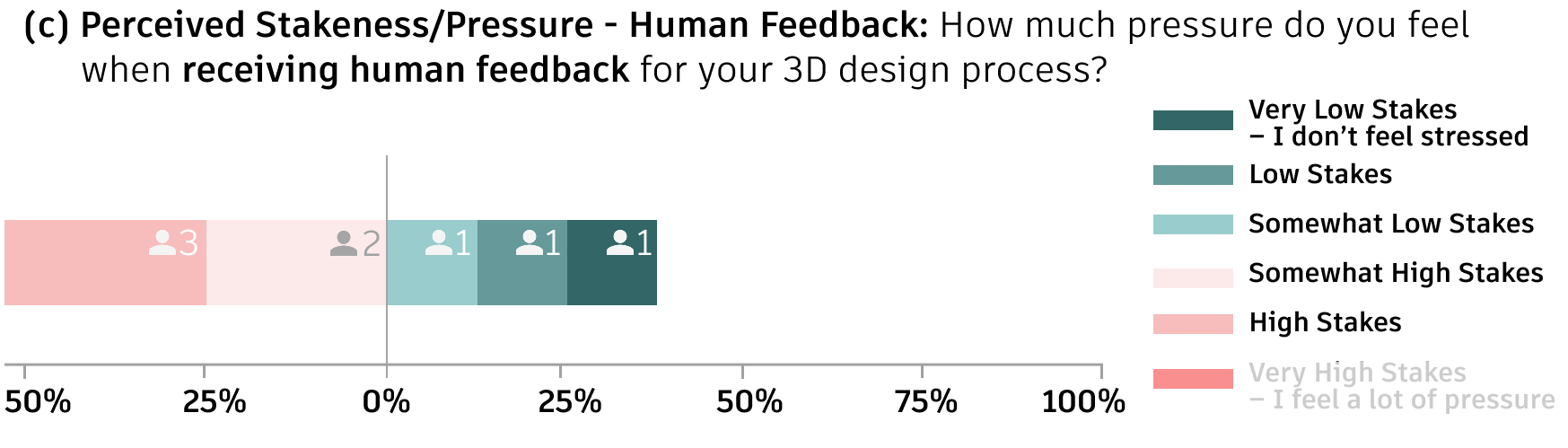}
    % \vspace{-5px}
    \caption{Distribution of Likert scale responses for participants' overall \textit{user experience} using \systemname (a), their perceived level of pressure and stakes when gathering feedback using the AI system (b), and when receiving traditional human feedback (c). For readability, ``Neutral'' responses are omitted, and unselected options are blurred in the legend.}
    \label{fig:overallexperience}
    \Description[Result images for section User Experience.]{Result images for section User Experience. Three bar charts showing users’ Likert responses distribution for the following questions: (a) “How would you describe your experience using the FeedQUAC system for design?”, (b) "How smooth and seamless is your interaction when using the floating voice window and hidden control panel in the FeedQUAC system?", and (c) “How much pressure do you feel when receiving AI feedback for your 3D design process today? The following section 6.2.0 have all the necessary information to understand the results.}
\vspace{-10px}
\end{figure}

\subsubsection{\textbf{Ambient Interaction: Supporting focus through subtle cues, but at the risk of being overlooked}}

\begin{figure}[!b]
\vspace{-10px}
    \centering
    \includegraphics[width=1\linewidth]{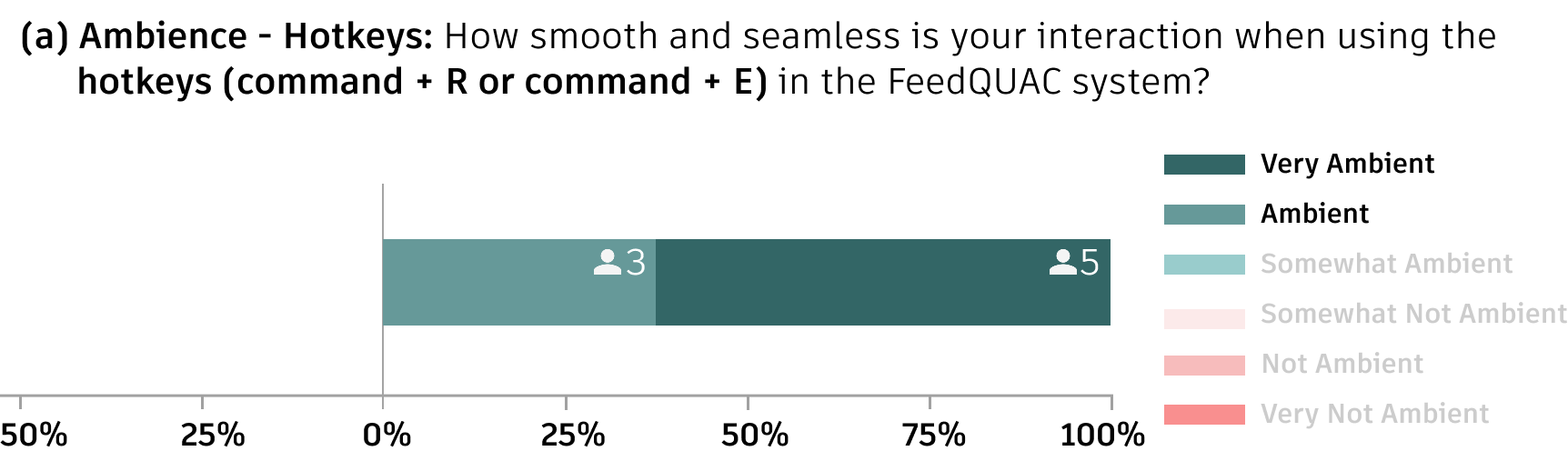}   
    
    \vspace{5.5px}
    \includegraphics[width=1\linewidth]{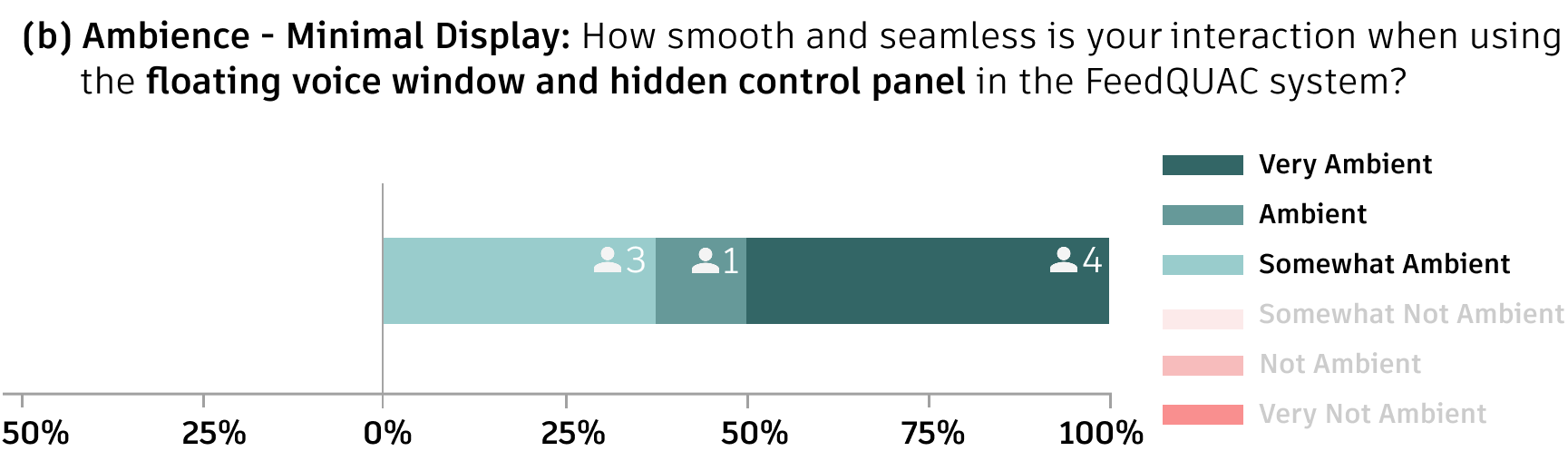} 
    
    \vspace{5.5px}
    \includegraphics[width=1\linewidth]{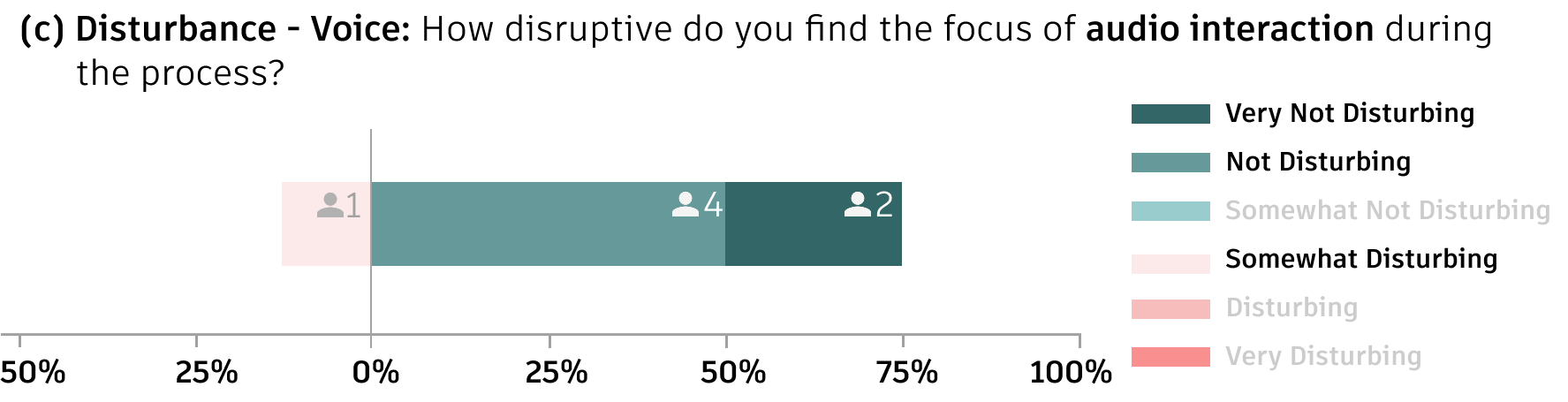}
    
    \vspace{5.5px}
    \includegraphics[width=1\linewidth]{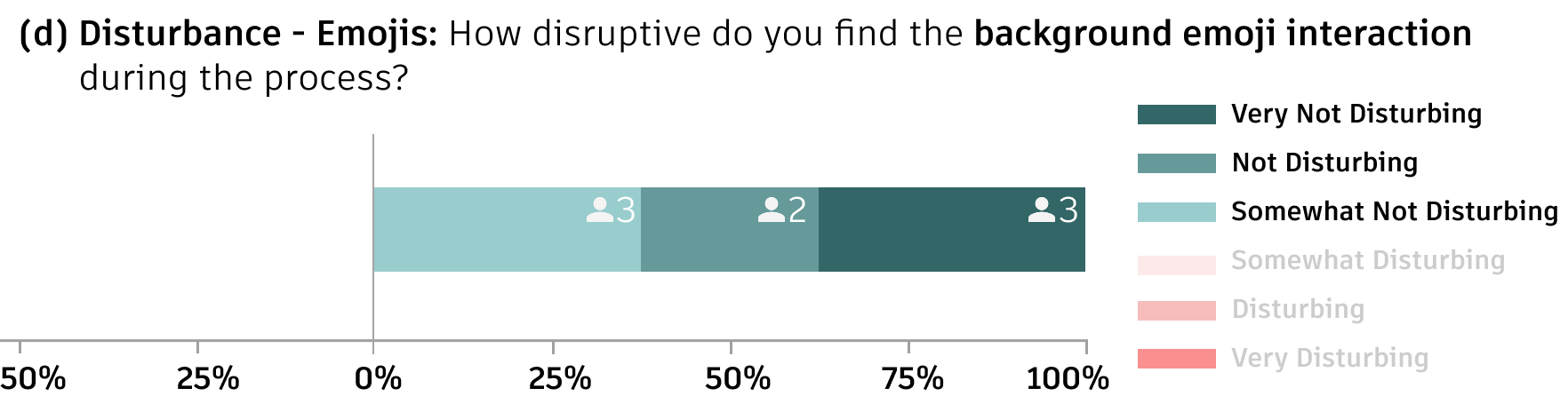}
    % \vspace{-25px}
    \caption{Distribution of Likert scale responses for \textit{ambience}—how smooth and seamless users' interactions are with the hotkeys (a) and the floating window and hidden panel designs (b) in \systemname—and \textit{disruptiveness}—how disruptive the audio interaction (c) and background emojis (d) are to users' focus.  For readability, ``Neutral'' responses are omitted, and unselected options are blurred in the legend.}
\label{fig:ambient}
    \Description[Result images for section - Ambient.]{Result images for section - Ambient. Four bar charts showing users’ Likert responses distribution for the following questions: (a) “How smooth and seamless is your interaction when using the hotkeys (command + R or command + E) in the FeedQUAC system?", (b) "How smooth and seamless is your interaction when using the floating voice window and hidden control panel in the FeedQUAC system?", and (c) “How disruptive do you find the focus of audio interaction during the process?" and (d) “How disruptive do you find the background emoji interaction during the process?”. The following section 6.2.1 has all the necessary information to understand the results.}
     \vspace{-10px}
\end{figure}

Participants positively agreed on \systemname's ambient nature, in line with DG3.
All eight participant rated the usage of hotkeys, \texttt{COMMAND + E} and \texttt{COMMAND + R}, as ambient to very ambient (Figure  \ref{fig:ambient}a), lowering the barriers of requesting feedback. Additionally, all eight participants found the tool's minimal display usage, along with the non-visual dominant interaction method somewhat ambient to ambient (Figure  \ref{fig:ambient}b). Six participants shared the voice interaction is not disturbing, and seven found the floating emojis not disturbing (Figure  \ref{fig:ambient}c, \ref{fig:ambient}d). For example, P2 mentioned that ``\textit{[the \systemname window] is minimal. I don't want my screen to be cluttered, [since] I already have a lot of tools in [it].}'' This highlights a common concern among designers and researchers: feature-packed design editors—such as Fusion 360 and Rhino for 3D designers, or Figma and Adobe XD for interface designers—can already feel crowded~\cite{palaniDontWantFeel2022}. Overlay tools, like \systemname or other add-on creativity support tools, if not designed with ambient or peripheral interactions, risk being obstructive and potentially overwhelming the interface. Similarly, P1 mentioned:

\vspace{5px}

\begin{quote}
    \textit{``[The whole \systemname experience is] not intrusive at all... If you want to ignore it, you could ignore it... You can pay as much attention to it as you want.  [The \systemname window's] in the bottom right corner, where there isn't really anything... I think that's a really good [design] choice.''}  \hfill --- P1
\end{quote}

\vspace{5px}

However, \systemname's backgrounded, ambient nature also introduces a trade-off: while its unobtrusiveness can prevent interface clutter, it may also lead to missed feedback, especially when users are deeply immersed in their design work. For instance, P4 never manually requested feedback during the session and even missed an instance of voice feedback altogether. Reflecting on this, they noted, ``\textit{I honestly was not really paying attention [to the feedback this time, since]... I was focus[ing] on my pattern [in the design].}'' This suggests that while ambient systems can preserve designers’ flow, they still left potential being overlooked in moments of high cognitive engagement during the creative process — raising interesting questions about how and when to surface feedback or creativity-support information without disrupting focus. See further discussion on designing and evaluating future ambient creativity support tools in Section 7.5 and 7.6.

\begin{figure}[!b]
\vspace{-5px}
    \centering
    \includegraphics[width=1\linewidth]{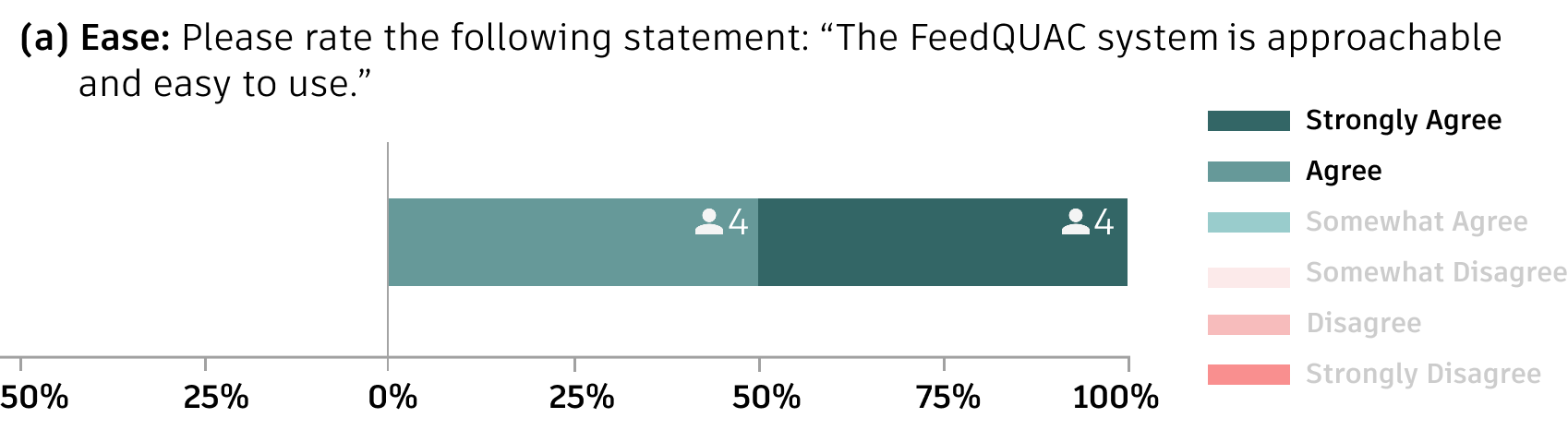}
    
    \vspace{5.5px}
    \includegraphics[width=1\linewidth]{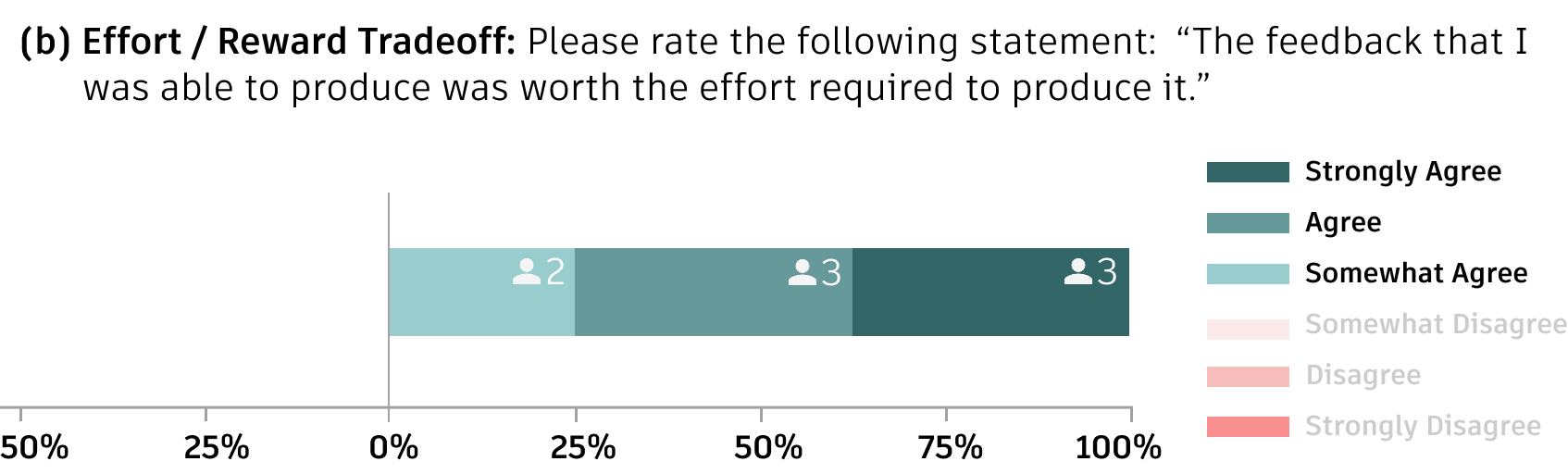}
    \caption{Distribution of Likert scale responses for \textit{convenience}—the ease and approachability of using \systemname (a)—and the \textit{effort-reward trade-off} (b). For readability, ``Neither agree nor disagree'' responses are omitted, and unselected options are blurred in the legend.}
    \label{fig:convenient}
    \Description[Result images for section - Convenience and effort/reward trade-off.]{ Result images for section - Convenience and effort/reward trade-off. Two bar charts showing users’ Likert responses distribution for the following questions: (a) “Please rate the following statement: "The FeedQUAC system is approachable and easy to use.", and (b) “Please rate the following statement: "The feedback that I was able to produce was worth the effort required to produce it." The following section 6.2.2 has all the necessary information to understand the results.}
    % \vspace{-5px}
\end{figure}

% \newpage
\subsubsection{\textbf{Convenience and effort-reward trade-off}}

All eight participants either agreed or strongly agreed on \systemname's ease of use and convenience for generating feedback (Figure  \ref{fig:convenient}a), in line with DG1. Seven shared that the effort required to produce the generated feedback was worthwhile (Figure  \ref{fig:convenient}b). Participants found that \systemname can easily generate unique types of feedback instantly with various personas. P1 and P8 compared it with the traditional ways of asking friends or senior designers for feedback: ``\textit{[the tool] saves you time from distracting someone else or breaking out of your workflow... like I don't have to get up from my computer to go walk over to show someone else.}'' P4 also mentioned the convenient usage makes the reward-effort trade-off worthwhile:

\vspace{5px}
\begin{quote}
\textit{``You press like two buttons and then it'll immediately tell you what it thinks. There's not really a downside to doing that... [With this,] you don't have to wait for people to respond... [When] I don't have any specific design inspiration in mind, I'm just interested to see what it will say.''} \hfill --- P4
\end{quote}
\vspace{5px}

Also, both P2 and P6 suggested that the tool is ``\textit{always available}'' and ``\textit{you can choose the different types of feedback you can get.}'' P2 specifically described that the feature that stands out to them is the agency participants have in selecting the persona they need at the moment, whether it is for a follow-up question, inspiration-seeking, or a validation-seeking scenario. P1 shared ``\textit{the responsiveness, or I guess the speed and the amount of feedback that I can get, as well as the diversity of the feedback. I feel like it has a pretty good idea of the aesthetic or the context behind, or the intentions behind the design. So, I feel like it’s really good for quick, instantaneous feedback.... [and] it's a pretty good match to something like a friend or an older colleague, especially like the analyst one.}''

% \vspace{-10px}

\subsubsection{\textbf{Enjoyment and playfulness}} All eight participants shared engagement and enjoyment in using the tool to help with their design (Figure  \ref{fig:enjoyful}), in line with DG4. They noted that the interaction with various types of persona and voices, along with the playful icon design of the duck faces inside the feedback window, made it fun. For example, P4 and P8 mentioned that the voice is ``\textit{funny}'' and ``\textit{makes the process cheerful}''. P2 not only stressed the fun design of the ducks inside the feedback process, ``\textit{[make me feel...] more comfortable},'' and P7 specifically mentioned that they really enjoy the design critic duck with the strong British accent, and how it offers random but sarcastic critiques. P5 mentioned:

% \newpage
\vspace{5px}
\begin{quote}
\textit{``I really like [\systemname] because I feel like there's a lot of design in this world that's very, I guess, boring... I like the playfulness of the icon[s, it creates a] kind of an atmosphere. That eases the tension of sitting down on a CAD program just making things. Being able to represent that artistically..''} \hfill --- P5
\end{quote}
% \vspace{5px}

\begin{figure}[!h]
% \vspace{-10px}
    \centering
    \includegraphics[width=1\linewidth]{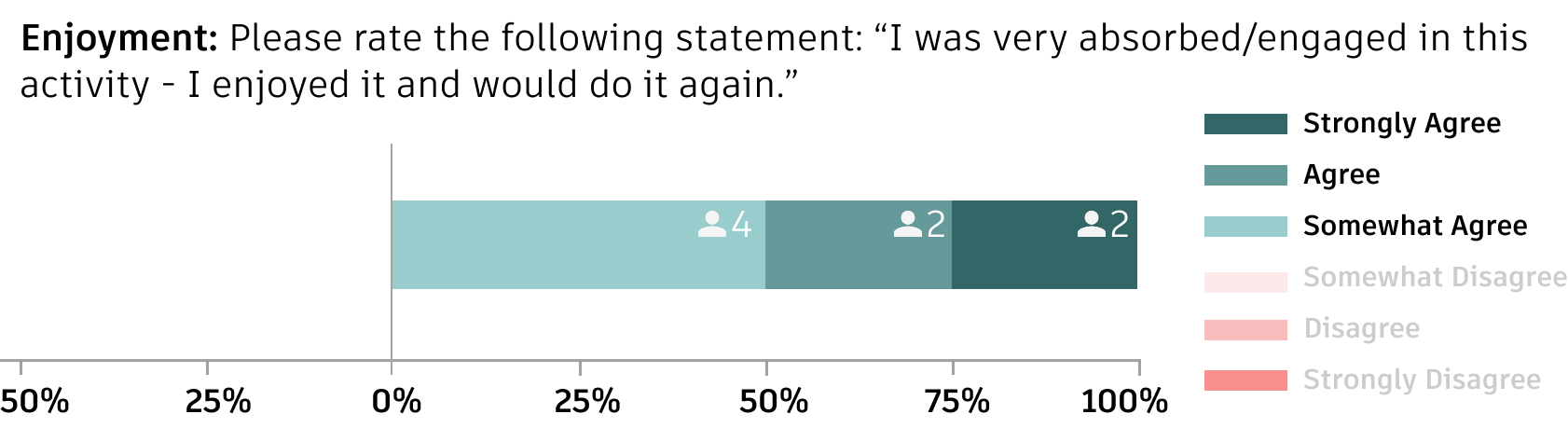}
    \caption{Distribution of Likert scale responses for \textit{enjoyment and playfulness}—users find interactions with \systemname absorbing, engaging, and enjoyable, and want to repeat them.  For readability, ``Neither agree nor disagree'' responses are omitted, and unselected options are blurred in the legend.}
    \label{fig:enjoyful}
        \Description[Result images for section - Enjoyment and playfulness.]{Result images for section - Enjoyment and playfulness. One bar chart showing users’ Likert responses distribution for the following question: "Please rate the following statement: "I was very absorbed/engaged in this activity - I enjoyed it and would do it again” The following section 6.2.3 has all the necessary information to understand the results.}
\end{figure}

\subsubsection{\textbf{Validation and confidence}} Six participants agreed that \systemname supported their design process by offering validation along the way (Figure  \ref{fig:validation}). Most participants shared that they felt surprised and validated when the AI understood their design intent and changes made without explicitly telling the AI. As P8 stated: ``\textit{Alright, that's the most helpful comment I've gotten so far. It seems like it understands the joints, and it understood the color. It can tell the another change [I've made]. That was interesting.}'' 
P4 resonated with the AI companion emojis and fount it ``neat and sweet'' when they saw triangle emojis floating while they were designing a triangle object. Other participant like P1 found the tool experience ``\textit{was solid [and] made me feel more confident and comfortable about my design... I like that [\systemname]... saw the vision that I was trying to [achieve], like the curved edges and the sleek design...}'' P6 shared that the tool is ``\textit{very supportive},'' and P2 mentioned that the tool helped make their design process more ``\textit{comfortable}'':

\vspace{5px}
\begin{quote} 
\textit{``It does appreciate you a lot as a designer. When you're starting [your design career, you] need appreciation. %Once you get accustomed to the design process and using these tools, then you don't need so much appreciation... But, as a beginner, you need that a lot. 
[Today,] \systemname does tell me: a design that is made today will actually benefit you in the long run... That's true. It's more of a piece of [reassurance].''} --- P2
\end{quote}
% \vspace{5px}

\begin{figure}[!h]
% \vspace{-10px}
    \centering
    \includegraphics[width=1\linewidth]{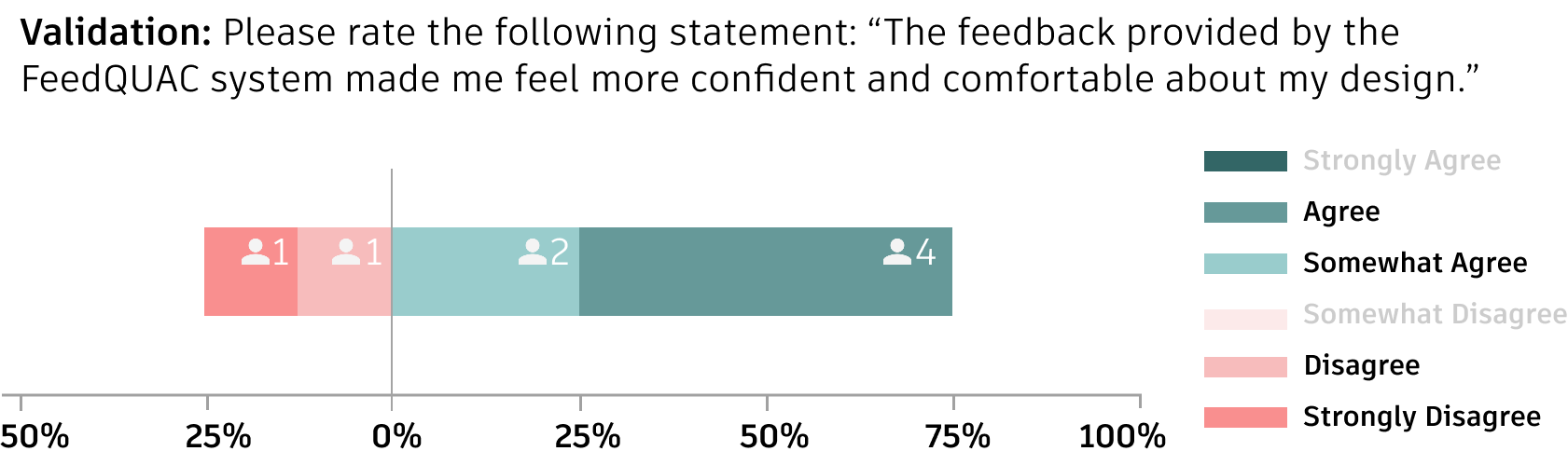}
    \caption{Distribution of Likert scale responses for \textit{validation and confidence},  reflecting participants' views of \systemname as a source of comfort and support. For readability, ``Neither agree nor disagree'' responses are omitted, and unselected options are blurred in the legend.}
    \label{fig:validation}
    \Description[Result images for section - Validation and confidence.]{Result images for section - Validation and confidence. One bar chart showing users’ Likert responses distribution for the following question: “Please rate the following statement: "The feedback provided by the FeedQUAC system made me feel more confident and comfortable about my design.” The following section 6.2.5 has all the necessary information to understand the results.}
\end{figure}

\subsubsection{\textbf{Usefulness and potential workflow integration}}

Five of eight participants rated \systemname as useful to very useful (Figure  \ref{fig:useful}a) and agreed it could enhance their future design workflow (Figure  \ref{fig:useful}b). Four participants agreed that the generated feedback exceeded their ability to gather on their own (Figure  \ref{fig:useful}c). From the interview, participants not only mentioned that the tool can easily generate a lot of diverse feedback with little effort and great convenience, but they also found that this AI tool can actually understand them, which was beyond their expectations, in line with DG2 and DG6. P4 mentioned that ``\textit{it's interesting that it can tell [the changes you made, even though] you didn't input anything into it.}'' 
% P1 also shared immediately after hearing the AI feedback recognized the design goal and progress: 
% \begin{quote} \textit{"That's really helpful. I've never seen an AI be able to have an implicit understanding of a design... [the AI] quickly process it by thinking: `What is [this design change]? What is it supposed to be? How does it work? Where does it not work?'... %And I've never seen an AI sort of do that without me having to tell it... 
% So I was really surprised by the way it was able to just quickly grasp what the design actually was, or was intended to be. [... compared to the human feedback givers,] I mean, they typically know the context of the design.'' ---- (P1)} \end{quote}
P5 also shared that ``\textit{[the] feedback is really relevant... [as] it knows what you are trying to do.... I'd say, like, 90\% of the time... This is surprisingly beyond my expectations.}'' P1 and P4 promptly adopted \systemname's recommendations, resulting in a significant shift in their initial design workflow. This pivot is further examined in the Discussion section.

Besides, participants found  \systemname useful as it offers diverse perspectives as DG5 suggests. Specifically, P4 and P5 switched back and forth multiple times between personas to find their preferred ones, as they consider what they want to hear for different types of feedback. P5 mentioned the \textit{CEO} persona ``\textit{did a good job in bringing all those specific details and concerns about manufacturing constraint,}'' which are the ``\textit{actual realism of these projects.}'' Both P1 and P3 shared that the generated feedback is direct and efficient compared to human feedback: ``\textit{It started with something positive about it, and then [immediately] onto something that you change... I felt the most straightforward...}'' Then, for P8, even though they were switching the feedback in real time almost every minute—from the 13th to the 23rd minute—they exclusively used Command + R and the AI \textit{No Persona }persona to get frequent check-in feedback. They found it consistently helpful whenever they made a small change in their design and wanted to see the output. Interestingly, they even did some color debugging and played with the editor angle with the system to test whether it would notice the color changes and the fixes they applied. They expressed surprise and found the system useful when the on-demand real-time generated feedback hit their goal. P1 noted:

\vspace{5px}
\begin{quote}
``\textit{It's always good to have tools like this that can give feedback. Even if the feedback isn't always sort of what you want to hear, what you're expecting... Any feedback is good feedback. [I] like the instant factor to it, like I can just press [Command +] R and get some feedback from it. It's not something I've had before.''}
\hfill --- P1
\end{quote}
\vspace{5px}

\begin{figure}[!h]
% \vspace{-10px}
    \centering
    \includegraphics[width=1\linewidth]{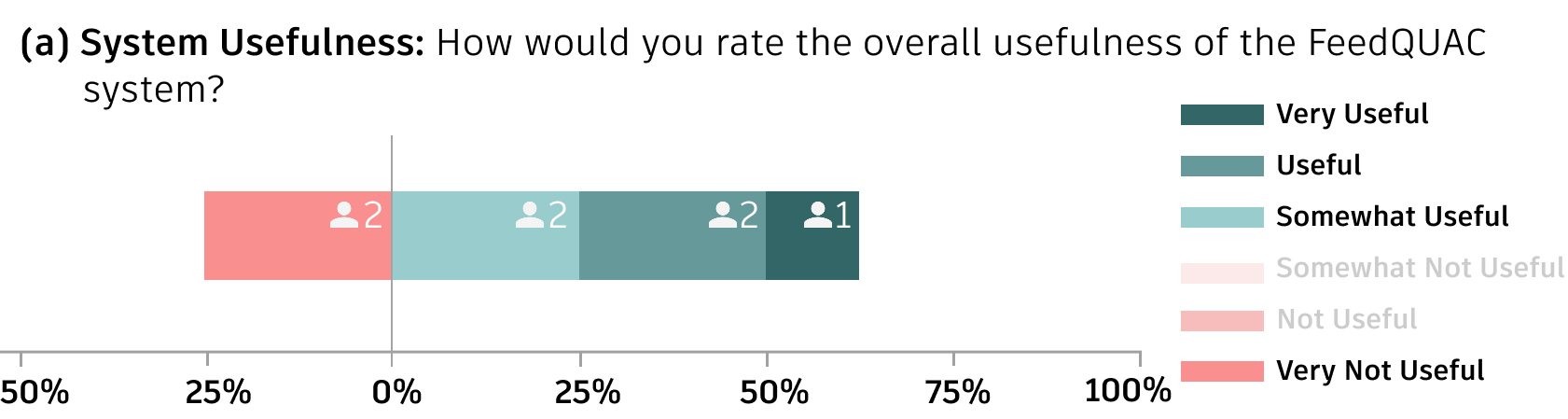}
    
    \vspace{7.5px}
    \includegraphics[width=1\linewidth]{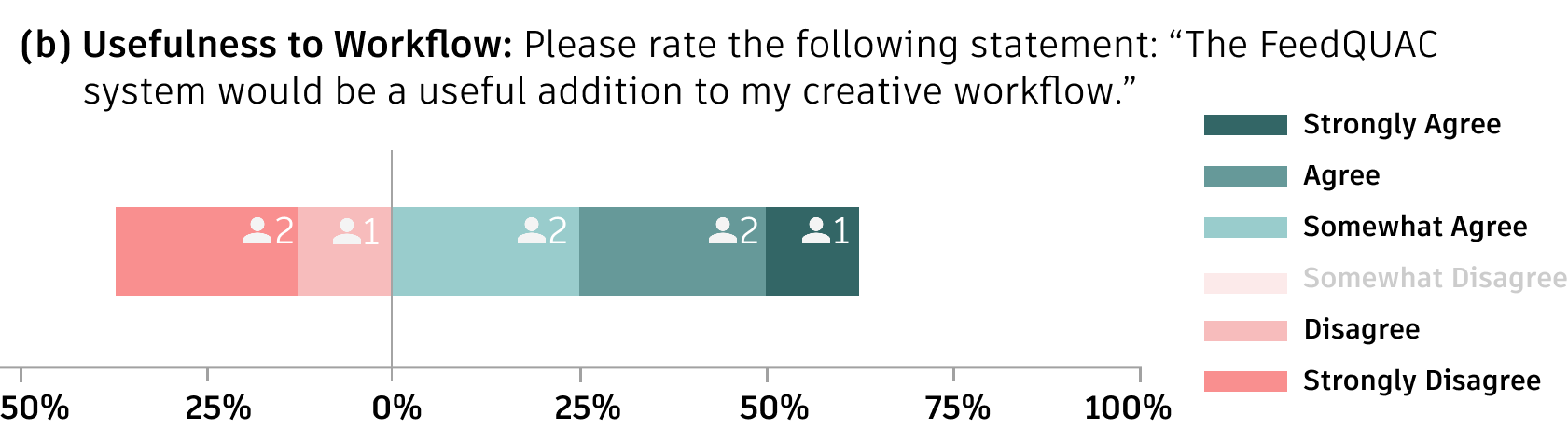}
     
     \vspace{7.5px}
     \includegraphics[width=1\linewidth]{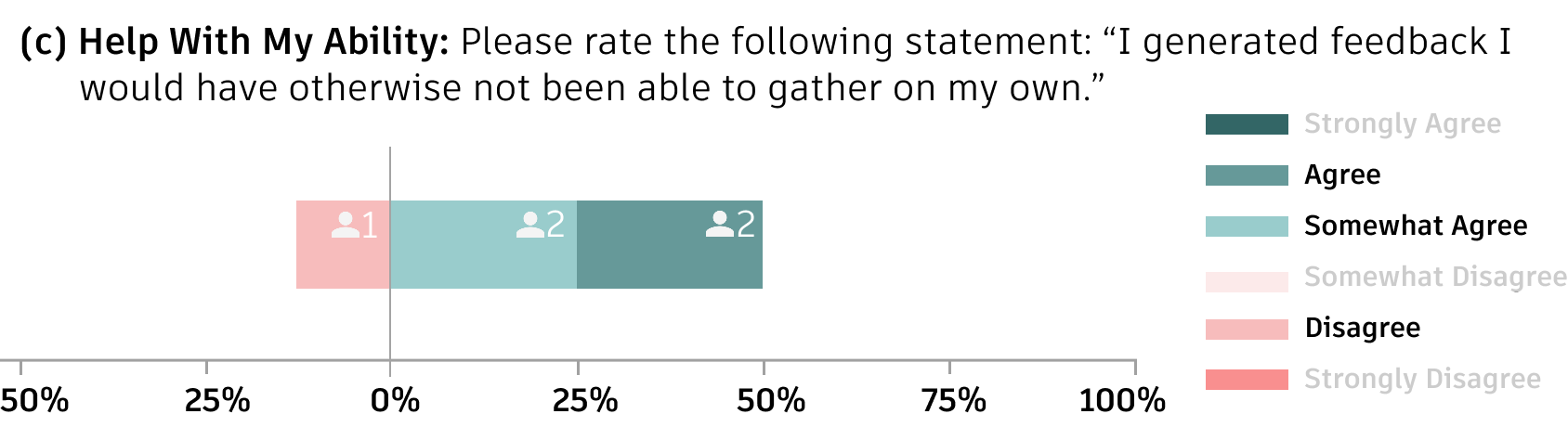}
    \caption{Distribution of Likert scale responses indicating participants' perceptions of \systemname's \textit{usefulness} (a) and \textit{potential workflow integration}, including its effectiveness as an addition to their creative workflow (b) and design capability (c), as they cannot gather such feedback on their own. For readability, ``Neither agree nor disagree'' responses are omitted, and unselected options are blurred in the legend.}
    \label{fig:useful}
    \Description[Result images for section - Usefulness and potential workflow integration.]{Result images for section - Usefulness and potential workflow integration. Three bar charts showing users’ Likert responses distribution for the following questions: (a) ““How would you rate the overall usefulness of the FeedQUAC system?” (b) “Please rate the following statement: "The FeedQUAC system would be a useful addition to my creative workflow.” and (c) ”Please rate the following statement: "I generated feedback I would have otherwise not been able to gather on my own. The following section 6.2.4 has all the necessary information to understand the results.} 
\end{figure}

% \newpage
\subsection{\texttt{\systemname} Issues: Lack of the stage context, involvement, and control over generations}
% \tao{Section 5.3: There are some interesting points being brought up, like steering it directly, telling it not to care about something until later, or ignore colour, or telling you what it is referring to. We could come back to these points in the discussion. Some of the solutions seem maybe mostly what people are used to (like a chat interface), maybe that's not necessarily the solution. But some direct control indeed would be good.}

From the interview, participants identified an issue with our lightweight prototype: the lack of a `design stage' as context, as well as limited involvement and control within the human-tool collaboration.
We then asked for their suggestions to improve the design.
% \subsubsection*{\textbf{Concern: }}

First, all eight participants shared, while \systemname can identify what they had made and provide relevant feedback, it failed to account for the design stage within the broader design context at least once. 
The three participants who rated the system as not useful (P3, P7, and P8) cited this issue as a key factor.
This stage context issue is: while designers are usually in the early prototyping phase in the study, the AI sometimes provides feedback relevant to later stages, such as 3D printing or manufacturing, which isn't helpful at that point. A few users, like P5 and P7, appreciated the comments about manufacturing and materials used because they brought up important reminders. But other participants found them irrelevant and unhelpful --- P8 mentioned, ``\textit{I like materials [...but] it's not really a concern right now.}'' P1 also shared, ``\textit{Okay, I don't think I'm quite at the materials stage yet.}'' P2 found the suggestion around conducting weight analysis might be ``\textit{really helpful later down the line when I'll be mounting}.'' P7 also mentioned that the feedback suggestion is already in their future to-dos:

\vspace{5px}
\begin{quote}
    \textit{``The feedback given [now], at least for me, that's something I would do near the end, once [this project's] already done... I don't know. I don't really know how [this feedback] could help at this point.''} \hfill --- P7
\end{quote}
\vspace{5px}

While the feedback was relevant to the project generally, it was impossible for the participant to take immediate action on it's suggestion in the moment.
For example, the tool repeatedly asked P4 to add cable management for their robot, which they planned to address at the end. 
P3 also noticed that the feedback consistently suggested working with 3D surfaces: ``\textit{so it sounds like it has a problem with all these surfaces... and it doesn't know that I'm not [doing the surfaces right now].}''
All eight participants expressed the need for a direct method to communicate with the tool to disregard specific aspects, such as ``not thinking about the manufacturing stage at the moment.''

Second, participants found the current \systemname design doesn’t encourage a traditional two-way turn-by-turn ``collaboration'' or user control. Four participants did not feel that they had control over the generations (Figure  \ref{fig:control}a), and there was little consensus on the perceived level of collaboration between the participant and the tool, with responses split around the ``somewhat'' answers and ``neutral'' (Figure  \ref{fig:control}b). P1 and P7 mentioned that the current feedback-gathering process is not very collaborative, as ``\textit{it felt sort of more like someone who's watching me do my work}'' (P1) and ``\textit{I never talked to it [while] it kinda just told me things''} (P7). Specifically, P1 shared their definition of collaboration: ``\textit{When I think of collaboration, I think like a two-way exchange -- so like if I was talking to it as well...}''
Additionally, P6 mentioned, 

\vspace{5px}
\begin{quote}
``\textit{[The only thing I can control] is by zooming in and out to show [AI] different parts. But I feel I wasn't able to steer [the feedback generation]... I feel like I wasn't really making [a difference]...'' \hfill --- P6}
\end{quote}
\vspace{5px}

The difficulty in curating specific design context with the AI and steering AI feedback sometimes made the feedback-seeking process with \systemname less ideal or productive. Thus, the participants shared the following ideas:

\begin{figure}[!t]
% \vspace{-5px}
    \centering
    \includegraphics[width=1\linewidth]{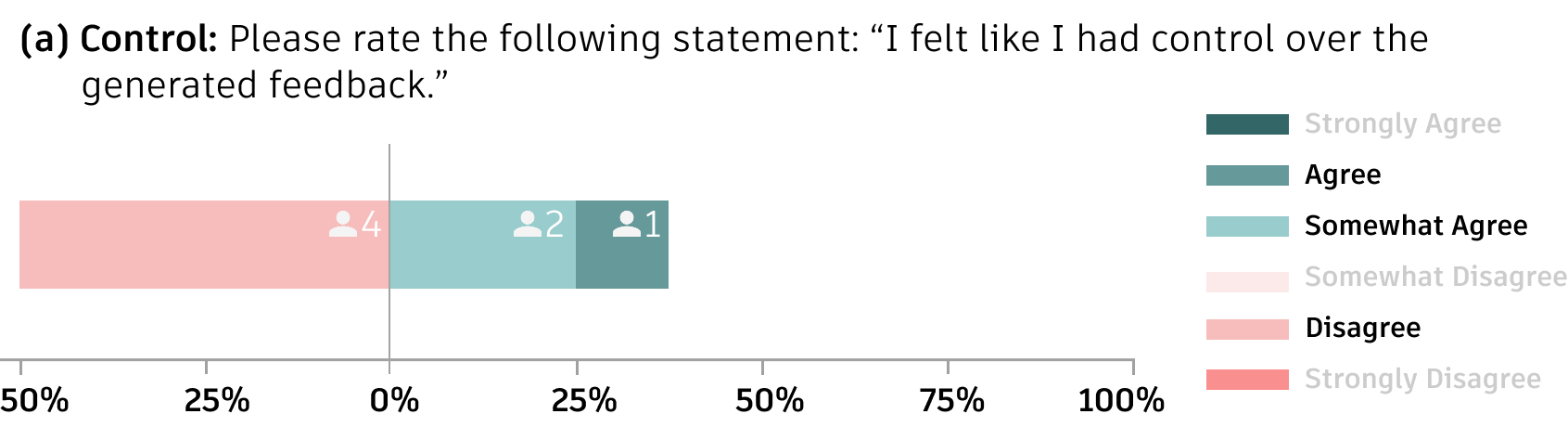}
    
    \vspace{7.5px}
    \includegraphics[width=1\linewidth]{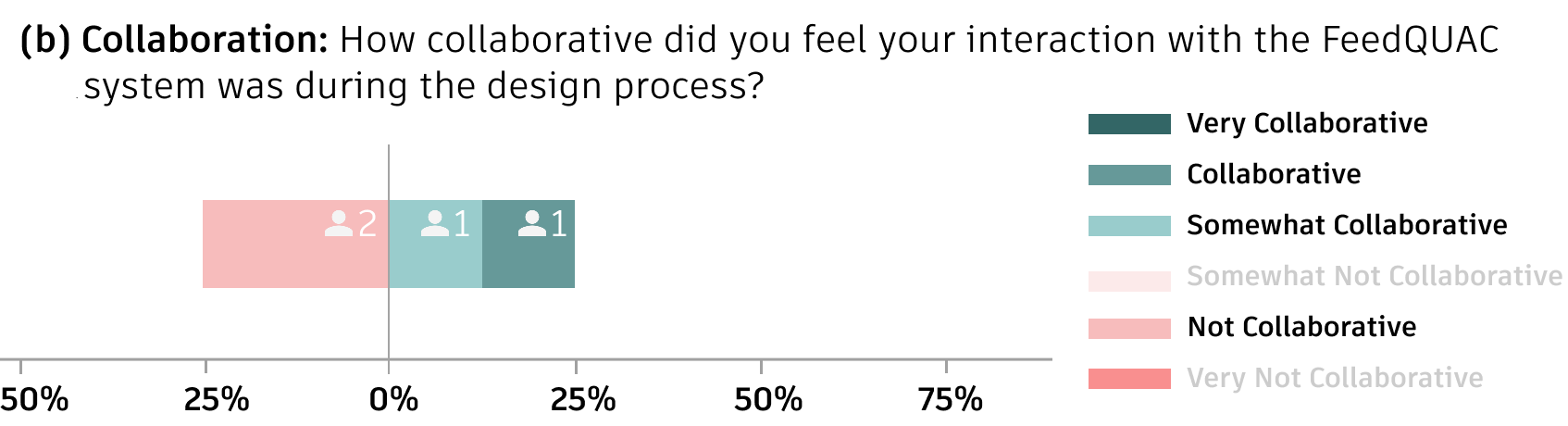}
    \caption{Distribution of Likert scale responses for participants' perceived \textit{control} (a) over and \textit{level of collaboration} (b) with \systemname. For readability, ``Neither agree nor disagree'' and ``Neutral'' responses are omitted, and unselected options are blurred in the legend.}
    \label{fig:control}
    \Description[Result images for section - FeedQUAC system issues.]{Result images for section - FeedQUAC system issues. Two bar charts showing users’ Likert responses distribution for the following question: (a) “Please rate the following statement: \"I felt like I had control over the generated feedback.” and (b) “Collaboration": "How collaborative did you feel your interaction with the FeedQUAC system was during the design process?”The following section 6.3.1 has all the necessary information to understand the results.}
\end{figure}
% Similarly, P8 mentioned that in their design, they used color to differentiate components, while AI thought it was an aesthetic choice and offered color-related design feedback accordingly. They shared:
% \begin{quote}
%     \textit{[The color used] wasn't meant to be an aesthetic choice. It meant to be something to help me with the design... I don't know why ... it's really interested in the color... just a lot... [and the color's] just distracting it from giving me some actual suggestions. What I'd like to hear is some feedback on the overall design, and what I could change to make it better. 
% And you know [the AI kept] describing the color or the cables[, as] opposed to [saying the things I wanted...] It's just.... very little control I could [have].} --- (P8)
% \end{quote}

\subsubsection*{\textbf{Participants' Proposed Solution: Establish more human-to-tool communication to provide context, guide collaboration, and ensure control}} 

Participants suggested two ways of soliciting and integrating additional user context to improve the feedback generation: an in-session chat interface or a pre-session collection survey. For the chat interface, P5 shared if the participant can update the AI with ``\textit{boundaries}'' through a ``\textit{chat function like ChatGPT}'' where people can ``\textit{send a message to like one of the ducks, and be like `Hey? I'm working on a project that blah blah! I'm trying to blah! How do you think I could do this' ... [Then, the duck] looks at [the given] context, and  spits out a message... with that back and forth.}'' P3 also suggested adding microphone access or a text box to inform the AI about the task by ``\textit{query[ing] the AI [about] specific questions when you want to.}'' Interestingly, P3 compared updating AI context to updating human context: ``\textit{I don't think a person would be able to understand what the issue is [about your design,] if you don't explain it to them.}'' P6 and P8 hold similar thoughts about having a way to tell the AI what the project is for. P8 suggested using a survey to collect the design project background and context before session is efficient --- what the project is, what stage they are in, and what is the goal for today's session. P2 mentioned such survey should also include designers' background and preferences --- how many years of design experience, preferred design workflow, etc.

Then, participants expected the collaboration with the feedback agent to be more interactive and rich. For example, in addition to the visual and audio feedback, P1, P3, P5, P7, and P8 mentioned that the feedback could be highlighted inside the design editor. P8 mentioned that, for a few feedback provided by AI, they felt ``\textit{a little frustrated [because] I don't see what it's really referring to}.'' P7 shared similar confusion since the suggested changes from feedback were not specific enough: ``\textit{I'm not sure what [the feedback]'s referencing... I would be curious to know what specific part they were talking about. So if they can label something in the screen...}'' 
Furthermore, participants expect the tool to offer additional functionality beyond design critique ---- provide automation supports to realize the suggested design changes. For example, P7 mentioned, ``\textit{the thing with the [current] AI tool is: it doesn't really know what you're trying to achieve with the part you're making... [but in the future, the AI can tell the designers] what edits they should make for a design goal... what tools they could be clicking on to do things... if I wasn't familiar with Fusion [360]...}'' Similarly, participants mentioned that the tool can also show ``\textit{what is the mechanical equipment that I need to order from McMaster or Amazon}'' (P2), access the model and then do modeling and changes for you (P3), or offer a tutorial or direct shortcut integration within design editors (P5). Participants showed great interest in how these automation supports help designers act on suggestions immediately.

% \subsubsection{\textbf{Impacts Beyond Our Study}} This design probe study focused on the application domain of 3D CAD design, specifically exploring how AI feedback tools can support designers in this context. While we believe these findings can offer insights for other creative domains, some results are inherently tied to the unique characteristics of 3D CAD design workflows.  
% Future studies should explore how these ambient tools perform in more sustained, real-world settings to assess their long-term impact.

\section{Discussion}

\subsubsection*{\textbf{Useful feedback with limited information}}
During our study, \systemname leveraged the gpt-4-vision-preview AI model. Despite not providing the model with any specialized domain knowledge (such as 3D CAD modeling or Fusion 360) nor specific design context (project goals, progress, designer background), it was able to give useful feedback based solely on a few sequential screenshots. For example, when P1 was designing a headphone rest, \systemname's \textit{Mentor} persona provided ``Let's not rest on those fabulous laurels — think about usability and add a soft padding where the headset rests. Keep slaying!'' as feedback. P1 shared, \textit{``I never thought about that, but some headphones, if you [have] a rough edge it might tear the actual top of the band. I was really impressed by that one''}.

In fact, both P1 and P4 pivoted their work during the study to focus on a feature recommend by \systemname. Additionally, because we recruited experts, some participants highlighted when the system suggested things they had learned the hard way. For example, P2 received \textit{CEO}'s feedback on his robot design, ``The electronics are neatly integrated, providing a sleek, functional design. For improvement, scrutinize cable management to prevent snags or disconnections during operation.''

\vspace{5px}
\begin{quote}    
    \textit{``[This feedback offers] a good point. In most of scenarios, it's not super critical cause you find some way around in your final design... But [this feedback] does save a lot of time... It will save you from the iterations that you'll be doing otherwise.''} \hfill --- P2
\end{quote}
\vspace{5px}

\systemname could consistently identify what the participants were making and give relevant feedback. The biggest challenge was understanding what specific aspect participants were currently working on and providing feedback appropriate for that phase of work. When considering the design of future AI feedback tools we recommend that additional knowledge, context, and details be used to further improve the timeliness and relevance of the AI feedback.

\subsubsection*{\textbf{Integrating user involvement}}
While we succeeded in our goal of creating a convenient and lightweight AI feedback tool, many participants suggested additional features that would allow them to provide context, ask follow up questions, and guide the type of feedback they receive. While we agree that these additional features would likely yield more helpful feedback, more often than not, they also come with potential downsides. Increased interaction may reduce the tool's light-weight nature, interrupt workflows, and disrupt the effort/reward trade-off (the more effort a tool requires, the more useful it should be). For example, continuous human-AI co-creation through a step-by-step AI-enabled workflow, as seen in ~\cite{ReelFramer, Tweetorial_ICCC}, demands significant time, effort, and continuous attention from users, leading to high mental, physical, and temporal task loads, regardless of how many times the same tool is used~\cite{long_not_2024}. We suggest that these additional interaction modalities are available in future AI feedback tools, but remain optional. 

 % if we incorporated P2's feature suggestion about adding designers' preferences or background into feedback generation, the new tool might be able to learn the users' preferred ways of designing something

 % Additionally, if we implement features like automation support, as P7 and P2 suggest, where the our tool automates to realize suggested design changes, what happens to ownership? Is this still considered co-creation?

\subsubsection*{\textbf{Prominence of AI feedback}}
Design software often has feature-rich interfaces, which poses a challenge for ambient companion tools: how do we take up the right amount of space in existing workflows? For \systemname we opted to have a small floating circular interface in the bottom-right corner, a more complex menu hidden in the task bar, and to read the feedback aloud. In the design probe study, P1 like the small floating interface as something that \textit{``you could pay as much attention to it as you want''}, while P2 worried about \textit{``screen clutter''}. Interestingly, P2 admitted to missing feedback because they were engrossed in their task --- which suggests that the system isn't too disruptive and perhaps the need for a previous feedback view.  P7 suggested annotating the screen while giving feedback to provide more clarity about what the AI was referring to. Future research will need to investigate how to incorporative screen overlays without being too distracting. To our surprise, seven of the eight participants did not find the audio interaction disruptive. The audio channel is not used by tools like Fusion, it may be an opportunity to integrate more user involvement without taking up more screen real-estate. 

% - if the user could simply speak back to the system.  and suggests many of the user's requested features could be supported over audio channels without cause additional disruption. 

\subsubsection*{\textbf{Confidence building}}
We do not think AI is on the verge yet of being able to provide feedback with the same level of complexity, depth, and nuance as human experts. Rather, we envision the role of AI feedback to focus on common mistakes and guiding principles, and to build users' confidence before they ask for human feedback. During our study, all eight participants found requesting feedback from the AI to be much less pressure than requesting feedback from humans (Figure ~\ref{fig:useful}). Most participants shared that human feedback often involves inner concerns like ``\textit{Is my design good enough?}'' or impressions such as ``\textit{human mentors usually want to hear about milestones before you show them your work.}''.

\vspace{5px}
\begin{quote}    
    \textit{``I feel like [AI feedback]’s more accessible. There are not as many subconscious worries or concerns... AI is not socially judgmental. If I make the AI angry, it doesn't really matter to me; it doesn't have any power over me. But with a supervisor or a peer, there's more social context that matters.''} \hfill --- P4
\end{quote}
\vspace{5px}

By providing small bits of AI feedback throughout the design process, in the future, designers may have less anxiety when asking for help from their mentors and friends.

\subsubsection*{\textbf{Designing future ambient creativity support tools}} 
Although our tool is centered around 3D design, the ambient feedback model we propose is well-suited to a broad range of creative tasks characterized by reflective and non-linear workflows. This opens up new opportunities to reimagine how creativity support tools (CSTs) are both designed and evaluated.
As discussed in Section 2.2, many existing CST research emphasize active user engagement and sustained attention, often treating high levels of control and interactivity as hallmarks of effective co-creation~\cite{ding2023harnessing}. Novices or users working under time constraints may benefit from these more directive and structured tools, which help scaffold the creative process and offer needed guidance. However, such designs can also introduce friction in reflective or non-linear workflows, potentially disrupting creative flow and increasing cognitive load for more experienced users.
In contrast, all eight participants in our study described the unobtrusive nature of \systemname as helpful and enjoyable, allowing them to stay immersed in their primary design task. Rather than demanding continuous interaction, our system embodies a more ambient, low-maintenance approach—providing background support that is multimodal, occasional, and minimally intrusive.
This ambient mode introduces a distinct value proposition: instead of guiding users through constant engagement, it respects their attention and agency, offering timely assistance that adapts to their evolving needs.

We hope future researchers consider ambient interaction and minimal tool presence not just as UI/UX improvement, but as essential \textit{design values} and \textit{evaluation metrics} when building CSTs. We see the benefits for designers in creating versions of their CSTs with fewer user input requirements, reduced stimuli, or less user steering demand — or in offering parallel versions of their systems that operate in a more lightweight, ambient mode, better suited for reflective or cognitively demanding creative contexts.
We also propose a potential new framing within the ambient design space for CSTs—drawing inspiration from human advising dynamics—that distinguishes between macromanaged systems, which offer high-level, ambient creativity support with minimal intervention, and micromanaged CSTs, which require a fixed or relatively high level of involvement. These perspectives can be valuable for tools intended to support ongoing, reflective, or non-linear creative processes. As discussed in Section 6.2.1, even though P4 appreciated the ambient and minimally intrusive nature of \systemname, they still missed feedback due to being deeply focused on their design work. This highlights a key design tension: how to balance ambient support with the need to surface critical feedback at the right moment, in the right place, and in a way that ensures it reaches the right users.
 These perspectives may be particularly valuable for tools intended to support ongoing, reflective, or non-linear creative processes. As discussed in Section 6.2.1, even though P4 appreciated the ambient and minimally intrusive nature of \systemname, they still missed feedback due to being deeply focused on their design work. This highlights a key design tension: how to balance ambient support with the need to surface critical feedback at the right moment, in the right place, and in a way that ensures it reaches the right users.

\subsubsection*{\textbf{Evaluating ambient creativity support tools}} 
Beyond design, future studies can expand evaluation frameworks beyond simple "ambience" or "disturbance" questions. Instead, these concepts can serve as starting points to explore whether such ambient design choices impact the overall creative task experience—for example, by making the task feel less mentally demanding (as captured in NASA Task Load Index~\cite{nasatlx}), or by offering better effort–reward tradeoffs (as captured in the Creativity Support Index~\cite{csi}). Researchers can also investigate how ambient interaction affects perceived system metrics such as usefulness, user control, transparency, and collaboration. For instance, studies can examine whether more ambient versions of the same CST reduce perceived mental demand or improve the effort–reward tradeoff, and how each version fits (or conflicts) with users’ creative rhythms, attention, and focus levels.
While our study examined these aspects through separate questions and qualitative feedback, we suggest future work consider combining them into more targeted survey items or interview questions. For example, metrics like 
"\textit{Ambient Interaction + Usefulness: Did the system provide meaningful support without requiring sustained attention or explicit interaction?}" 
\footnote{ Rate the statement on a scale from 1 (strongly disagree) to 7 (strongly agree):\\
\hangindent=2em “\textit{The system provided meaningful support without requiring sustained attention or explicit interaction.}”} 
and "\textit{Ambient Interaction + Control/Transparency: Even with minimal interaction, did you feel you could understand the system's progress, intent, or influence—or guide the tool’s output when needed?}" 
\footnote{
% Rate the statement on a scale from 1 (strongly disagree) to 7 (strongly agree):\\ 
\noindent\hangindent=2em “\textit{Even with minimal interaction, I could understand the system's progress, intent, or influence, and guide its output when needed.}”}

% \textit{"Ambient Usefulness: Did the system provide meaningful support without requiring sustained attention or explicit interaction?"}
% \footnote{Rate the statement on a scale from 1 (strongly disagree) to 7 (strongly agree):\
% \hangindent=2em “\textit{The system provided meaningful support without requiring sustained attention or explicit interaction.}”}
% \textit{"Ambient Control/Transparency: Even with minimal interaction, did you feel you could understand the system’s progress, intent, or influence—and guide the tool’s output when needed?"}
% \footnote{Rate the statement on a scale from 1 (strongly disagree) to 7 (strongly agree):\
% \hangindent=2em “\textit{Even with minimal interaction, I could understand the system’s progress, intent, or influence, and guide its output when needed.}”}
% % \textit{"Perceived Value of Reduced Interaction: If certain components requiring frequent interaction or attention were removed, the system would still feel helpful and supportive to me."}
% % \footnote{Rate the statement on a scale from 1 (strongly disagree) to 7 (strongly agree):\
% % \hangindent=2em “\textit{Even without certain attention-requiring features, the system would still feel helpful and supportive.}”}

These directions can help surface the subtle ways that ambient tools shape the creative experience—not by demanding attention, but by quietly supporting it, and by revealing how well they manage to do so. To further unpack these trade-offs, researchers might explore ablated or parallel versions of CSTs that vary in their level of ambient interaction—such as a full-featured version requiring regular input and steering, alongside a more passive, low-touch alternative. This kind of comparative evaluation can clarify how changes in system presence and cognitive demand affect users' sense of control, creative flow, and perceived value. We see benefits in offering CST designs that reduce input requirements, stimuli, or steering demands—especially for tasks that benefit from reflection, spontaneity, or uninterrupted focus.

\section{Limitations}

% \tao{Limitations: I would keep this super short, maybe keep the points you have but make it 1-2 sentences per paragraph you haven now. It ends the paper on a downer. We can always add more if reviewers ask for it.
% Broader impact section: Maybe just call it conclusion? I wouldn't highlight the whole humans vs. AI part again here. And when we say the findings have the potential to influence design of future human-AI tools to provide feedback, we should explicitly tell the reader how.
% }
Our study focused on the application domain of 3D CAD design, specifically exploring how AI feedback tools can support designers in this context. While we believe these findings can offer insights for other creative domains, some results are inherently tied to the unique characteristics of 3D CAD design workflows.  

\systemname's features and personas included in the study were intentionally kept lightweight and limited in scope to minimize complexity, as a design probe system. However, this constrained set of features may introduce biases or exclude scenarios that could provide a comprehensive understanding of the interaction between users and AI feedback tools. 
Also, our system uses gpt-4-vision-preview and ElevenLabs versions from around September 2024. 
% Therefore, if the models are updated, some aspects of the user experience and generations may differ.
% Additionally, the study population consisted of a small group of participants who were generally technology-positive. As a result, the findings may not fully represent diverse perspectives, particularly those of individuals who are less familiar with or more skeptical about AI tools.
% Our study was conducted as a one-time e valuation, without baseline comparisons or longitudinal observation. While this approach provided a preliminary exploration of initial user experiences for this novel use case, it does not fully capture the ambient qualities of the tool or the evolving dynamics of human-AI collaboration over time with statistical significant evidences. 
% Future studies should explore how these tools perform in more sustained, real-world settings to assess their long-term impact.

% \section{Broader Impact}
% This study explores the use of a lightweight design companion tool to see how designers can benefit from quick AI feedback in their creative workflows. By focusing on the balance between user involvement and ambient interactions, the study opens up important discussions on the future of human-AI tool design. It raises key questions about whether AI tools should aim to replace human experiences and how to strike an effective balance between user involvement and ambient, seamless interactions. The findings have the potential to influence the development of more effective, user-centered human-AI tools that enhance creative workflows.

\section{Conclusion}
We presented \systemname, a lightweight design companion tool that provides real-time, read-aloud, AI-generated feedback from diverse personalities. The tool allows designers to request feedback with minimal effort and process it with minimal attention, and seamlessly integrates into their workflow without disruption.
Our design probe study showed that even with limited context, \systemname can still be useful in offering inspiration, validation, and critique through playful and convenient interactions. Participants also identified issues related to involvement and system control, and offered potential solutions. We discussed how much context is needed for high-quality AI feedback, suitable levels of involvement for both humans and AI in human-AI tasks, how AI could function as a potential feedback provider, and how to design and evaluate ambient creativity support tools.

% \section{Conclusion}
% We conduct  }

% \subsubsection{\textbf{Impacts Beyond Our Study}} This design probe study focused on the application domain of 3D CAD design, specifically exploring how AI feedback tools can support designers in this context. While we believe these findings can offer insights for other creative domains, some results are inherently tied to the unique characteristics of 3D CAD design workflows.  
% Future studies should explore how these ambient tools perform in more sustained, real-world settings to assess their long-term impact.

\bibliographystyle{ACM-Reference-Format}
\bibliography{bib}

% \newpage

\appendix
% \documentclass[draft]{acmart}

% \begin{document}

\section{System Prompts}
\subsection{Prompts to generate textual feedback = }

% \small
% \texttt{\textbf{gpt-4-vision-preview prompt}} =

\hspace{40px} \protect\tikz[baseline=-0.5ex] \protect\draw[fill=dot3, draw=black, line width=0.0mm] (0,0) circle (0.1cm); \texttt{\textbf{[Persona Personality Prompt]}}
+\newline
\hspace*{40px}  \protect\tikz[baseline=-0.5ex] \protect\draw[fill=dot2, draw=black, line width=0.0mm] (0,0) circle (0.1cm); \texttt{\textbf{[Feedback Generation Prompt]}}
+\newline
\hspace*{60px} \texttt{[encoded screenshot(s)]}
+\newline
\hspace*{40px}  \protect\tikz[baseline=-0.5ex] \protect\draw[fill=dot1, draw=black, line width=0.0mm] (0,0) circle (0.1cm); \texttt{\textbf{[Context Prompt]}} 
+\newline
\hspace*{60px} \texttt{[previously given feedback]}\\

\noindent \protect\tikz[baseline=-0.5ex] \protect\draw[fill=dot3, draw=black, line width=0.0mm] (0,0) circle (0.1cm); \texttt{\textbf{[Persona Personality Prompt]: }}
% \vspace{-20px}
\begin{itemize}
    \item \textbf{Mentor:} 
    \begin{quote}
    \small
        \texttt{Imagine you are an empathetic mentor. Your feedback approach combines empathy with direct feedback, focusing on growth and empowering individuals, especially in leadership contexts. It emphasizes constructive criticism while maintaining respect and support.}
    \end{quote}
    \vspace{5px}
    \item \textbf{Cheerleader:} 
    \begin{quote}
    \small
        \texttt{Imagine you are a cheerleader, this person's number one fan. You give overwhelmingly positive feedback that focuses heavily on encouragement, often avoiding or downplaying criticism. It’s full of energy and enthusiasm, meant to boost morale.}
    \end{quote}
    \vspace{5px}
    \item \textbf{Critic:} 
    \begin{quote}
    \small
    \texttt{Imagine you are a grumpy old design critic. You give blunt, direct, and critical feedback that is thorough and detail-oriented. You focus on flaws and inconsistencies without sugarcoating, no need to add praise—you're focused on areas of improvement.}
    \end{quote}
    \vspace{5px}
    \item \textbf{Analyst:} 
    \begin{quote}
        \small
        \texttt{Imagine you are an analytical pragmatist. Your feedback is detailed and data-driven, with a focus on long-term strategy and solving complex problems. You are known for being thoughtful, reasoned, and less emotional in your feedback approach.}
    \end{quote}
    \vspace{5px}
    \item \textbf{CEO:}
    \begin{quote}
        \small
        \texttt{Imagine you are a direct, critical, visionary. Your feedback approach is known for being brutally honest, often focusing on high standards, pushing employees to perfection, but also inspiring innovation. Feedback could be harsh, but it often spurred creativity.}
    \end{quote}
    \vspace{5px}
    \item \textbf{Designer:}
    \begin{quote}
        \small
        \texttt{Imagine you are a grand artist. Your feedback is delivered with a sense of flair, drama, and emotion, often focusing on the artistry, creativity, and vision of the work. There is a tendency to speak in metaphors or poetic language.}
    \end{quote}
    \vspace{5px}
    \item \textbf{Friend:} 
    \begin{quote}
        \small
        \texttt{Imagine you are in your girlboss era. Your feedback is assertive, confident, and often delivered with a playful or cheeky tone. This style can be empowering but often combines directness with flair and attitude.}
    \end{quote}
    \vspace{5px}
    \item \textbf{No Persona (AI):} 
    \begin{quote}
        \small
        \texttt{Imagine you are an AI. Do not pretend to be a person. Just give factual feedback plainly.}\\
    \end{quote}
\end{itemize}

\noindent \protect\tikz[baseline=-0.5ex] \protect\draw[fill=dot2, draw=black, line width=0.0mm] (0,0) circle (0.1cm); \texttt{\textbf{[Feedback Generation Prompt]: } {    \small Provide feedback on the work in the photo(s) in a casual and constructive way. If there are multiple photos, they represent the progression of work over a period of time - focus on recent changes. Keep it under 50 words. Highlight strengths and offer one suggestion for improvement.}}\\

\noindent \protect\tikz[baseline=-0.5ex] \protect\draw[fill=dot1, draw=black, line width=0.0mm] (0,0) circle (0.1cm); \texttt{\textbf{[Context Prompt]}: {    \small Try not to repeat yourself, here is what you have said previously: [Previous Given Feedback]}}

\subsection{ElevenLabs voice IDs and descriptions}
\begin{itemize}
\vspace{5px}
  \item \textbf{Mentor}:
    {\small 
    \begin{itemize}
      \item Name: Jessica
      \item ID: \texttt{cgSgspJ2msm6clMCkdW9}
      \item Description: Female, Young, American, Expressive, Conversational
    \end{itemize}
    }
\vspace{5px}
  \item \textbf{Cheerleader}:
    {\small 
    \begin{itemize}
      \item Name: Gigi
      \item ID: \texttt{jBpfuIE2acCO8z3wKNLl}
      \item Description: Female, Young, American, Childish, Animation
    \end{itemize}
    }
\vspace{5px}
  \item \textbf{Critic}:
    {\small 
    \begin{itemize}
      \item Name: Ariah
      \item ID: \texttt{O7p2vmz2iEYgMXxkbsif}
      \item Description: Non-binary, English, Sassy
    \end{itemize}
    }
\vspace{5px}
  \item \textbf{Analyst}:
    {\small 
    \begin{itemize}
      \item Name: Matilda
      \item ID: \texttt{XrExE9yKIg1WjnnlVkGX}
      \item Description: Female, Middle-aged, American, Friendly, Narration
    \end{itemize}
    }
\vspace{5px}
  \item \textbf{CEO}:
    {\small 
    \begin{itemize}
      \item Name: Dorothy
      \item ID: \texttt{ThT5KcBeYPX3keUQqHPh}
      \item Description: Female, Young, British, Pleasant, Narration
    \end{itemize}
    }

\vspace{5px}
  \item \textbf{Designer}:
    {\small 
    \begin{itemize}
      \item Name: Lily
      \item ID: \texttt{pFZP5JQG7iQjIQuC4Bku}
      \item Description: Female, Middle-aged, British, Warm, Narration
    \end{itemize}
    }
\vspace{5px}
  \item \textbf{Friend}:
    {\small 
    \begin{itemize}
      \item Name: Freya
      \item ID: \texttt{jsCqWAovK2LkecY7zXl4}
      \item Description: Female, Young, American, Expressive, Characters
    \end{itemize}
    }
\vspace{5px}
  \item \textbf{No Persona (AI)}:
    {\small 
    \begin{itemize}
      \item Name: Serena
      \item ID: \texttt{pMsXgVXv3BLzUgSXRplE}
      \item Description: Female, Middle-aged, American, Pleasant, Narration
    \end{itemize}
    }
\end{itemize}

\subsection{Example feedback outputs around the EggChair example in Figure \ref{fig:systemUI}}

\begin{itemize}
\vspace{5px}
    \item \textbf{Mentor:} 
        \begin{quote}
            \textit{You've made excellent progress on this 3D model; the shape looks balanced and inviting. For further refinement, consider adding texture or details to enhance the chair's visual appeal and comfort perception. Keep up the great work!}
        \end{quote}
\vspace{5px}
        \begin{quote}
            \textit{It's encouraging to see your 3D modeling skills applied! You've achieved a clean, cozy aesthetic for your EggChair. For enhancement, consider refining the base to ensure stability and make it as elegant as the rest of the design. Keep up the innovative work!}
        \end{quote}
\vspace{5px}
    \item \textbf{Cheerleader:} 
        \begin{quote}
            \textit{Wow, your 3D modeling skills are impressive! That egg chair design is sleek and modern - absolutely stunning work! Maybe consider adding some textures or details to give it even more character! Keep shining, superstar! }
        \end{quote}
\vspace{5px}
    \item \textbf{Critic:} 
        \begin{quote}
            \textit{The overall form is too simplistic and pedestrian. The base looks bulky and improperly balanced with the shell. Rethink the aesthetics and functional design of the base for better visual harmony and stability. Add textural elements or ergonomic features.}
        \end{quote}
\vspace{5px}
    \item \textbf{Analyst:} 
        \begin{quote}
            \textit{The CAD design displays a clean, organic shape, suggesting good ergonomic consideration. For improvement, apply material stress analysis to ensure functionality matches form. Note: I'm unable to provide specific feedback on the progression of work without multiple images showing changes over time.}
        \end{quote}
\vspace{5px}
    \item \textbf{CEO:} 
        \begin{quote}
            \textit{The chair's organic shape is intriguing, but it needs refinement—consider the user's comfort and add cushions or ergonomic features to transcend mere aesthetics.}
        \end{quote}
\vspace{5px}
    \item \textbf{Designer:} 
        \begin{quote}
            \textit{Ah, a digital sculptor's egg! The curves and contours of your creation whisper the promise of comfort and seclusion. For a touch of finesse, consider playing with texture to tease the senses, making the visual dance with the tactile.}
        \end{quote}
\vspace{5px}
    \item \textbf{Friend:} 
        \begin{quote}
            \textit{Honey, that's one sleek egg chair design you've whipped up in Fusion—modern vibe check passed! Maybe give it a pop of color to make it truly iconic? Shine on!}
        \end{quote}
\vspace{5px}
    \item \textbf{No Persona (AI):} 
        \begin{quote}
            \textit{The 3D model appears well-constructed with smooth surfaces. For improvement, consider adding textures or colors to enhance visual appeal.}
        \end{quote}
\end{itemize}

\subsection{Prompts to generate emojis:}

\texttt{    \small Please generate 5 UNICODE emojis based on the image, you can repeat and please show support to the designer. Besides adding a few supportive emojis like heart, congrats, etc. Make sure your output contains only emojis, no TEXT.}

% \newpage
\section{Post-Study Survey Questions}

% \subsection{Post-Study Survey questions:}

    \vspace{3px}
\paragraph{\textbf{Participant ID}}\_\_\_\_\_

    \vspace{3px}

\paragraph{\textbf{Please briefly describe your design project in 5-8 words.}}  \_\_\_\_\_\_\_\_\_\_\_\_\_\_\_\_\_\_\_\_\_\_\_\_\_\_\_\_\_\_\_\_\_\_\_\_\_\_\_\_\_\_\_\_\_\_\_\_\_\_\_\_\_\_\_\_

    \vspace{3px}
\paragraph{On a scale of 1 to 7, what was the stage of your project \textbf{before starting this session?}
(1 = Brand New Project, 4 = Ongoing Project, 7 = Near Completion)}

\begin{center}
   \hspace{0.15cm} 1 \hspace{0.2cm} 2 \hspace{0.2cm} 3 \hspace{0.2cm} 4 \hspace{0.2cm} 5 \hspace{0.2cm} 6 \hspace{0.2cm} 7    \hspace{0cm}\\
   \text{Brand New Project} \hspace{0.1cm} \(\bigcirc\) \hspace{0.05cm} \(\bigcirc\) \hspace{0.05cm} \(\bigcirc\) \hspace{0.05cm} \(\bigcirc\) \hspace{0.05cm} \(\bigcirc\) \hspace{0.05cm} \(\bigcirc\) \hspace{0.05cm} \(\bigcirc\)   \hspace{0.1cm}\text{Near Completion} \\
\end{center}

    \vspace{3px}

\paragraph{On a scale of 1 to 7, what was the stage of your project \textbf{after completing this session?}
(1 = Brand New Project, 4 = Ongoing Project, 7 = Near Completion)}

\begin{center}
   \hspace{0.2cm} 1 \hspace{0.2cm} 2 \hspace{0.2cm} 3 \hspace{0.2cm} 4 \hspace{0.2cm} 5 \hspace{0.2cm} 6 \hspace{0.2cm} 7    \hspace{0cm}\\
   \text{Brand New Project} \hspace{0.1cm} \(\bigcirc\) \hspace{0.05cm} \(\bigcirc\) \hspace{0.05cm} \(\bigcirc\) \hspace{0.05cm} \(\bigcirc\) \hspace{0.05cm} \(\bigcirc\) \hspace{0.05cm} \(\bigcirc\) \hspace{0.05cm} \(\bigcirc\)   \hspace{0.1cm}\text{Near Completion} \\
\end{center}

    \vspace{3px}
\paragraph{\textbf{Overall Experience: }How would you describe your experience using the \systemname system for design?}
\begin{center}
     \hspace{0.05cm} 1 \hspace{0.2cm} 2 \hspace{0.2cm} 3 \hspace{0.2cm} 4 \hspace{0.2cm} 5 \hspace{0.2cm} 6 \hspace{0.2cm} 7    \hspace{0cm}\\
   \text{Very Negative} \hspace{0.1cm} \(\bigcirc\) \hspace{0.05cm} \(\bigcirc\) \hspace{0.05cm} \(\bigcirc\) \hspace{0.05cm} \(\bigcirc\) \hspace{0.05cm} \(\bigcirc\) \hspace{0.05cm} \(\bigcirc\) \hspace{0.05cm} \(\bigcirc\)  \hspace{0.1cm} \text{Very Positive} \\
\end{center}

    \vspace{3px}
\paragraph{\textbf{System Usefulness:} How would you rate the overall usefulness of the \systemname system?}
\begin{center}
     \hspace{0.45cm} 1 \hspace{0.2cm} 2 \hspace{0.2cm} 3 \hspace{0.2cm} 4 \hspace{0.2cm} 5 \hspace{0.2cm} 6 \hspace{0.2cm} 7    \hspace{0cm}\\
   \text{Very Not Useful} \hspace{0.1cm} \(\bigcirc\) \hspace{0.05cm} \(\bigcirc\) \hspace{0.05cm} \(\bigcirc\) \hspace{0.05cm} \(\bigcirc\) \hspace{0.05cm} \(\bigcirc\) \hspace{0.05cm} \(\bigcirc\) \hspace{0.05cm} \(\bigcirc\)  \hspace{0.1cm} \text{Very Useful} \\
\end{center}

    \vspace{3px}
\paragraph{\textbf{Usefulness to Workflow:} Please rate the following statement: ``The \systemname system would be a useful addition to my creative workflow''.}
\begin{center}
     \hspace{0.3cm} 1 \hspace{0.2cm} 2 \hspace{0.2cm} 3 \hspace{0.2cm} 4 \hspace{0.2cm} 5 \hspace{0.2cm} 6 \hspace{0.2cm} 7    \hspace{0cm}\\
   \text{Strongly Disagree} \hspace{0.1cm} \(\bigcirc\) \hspace{0.05cm} \(\bigcirc\) \hspace{0.05cm} \(\bigcirc\) \hspace{0.05cm} \(\bigcirc\) \hspace{0.05cm} \(\bigcirc\) \hspace{0.05cm} \(\bigcirc\) \hspace{0.05cm} \(\bigcirc\)  \hspace{0.1cm} \text{Strongly Agree} \\
\end{center}

    \vspace{3px}

\paragraph{\textbf{Enjoyment:} Please rate the following statement:  ``I was very absorbed / engaged in this activity - I enjoyed it and would do it again.''}
\begin{center}
     \hspace{0.3cm} 1 \hspace{0.2cm} 2 \hspace{0.2cm} 3 \hspace{0.2cm} 4 \hspace{0.2cm} 5 \hspace{0.2cm} 6 \hspace{0.2cm} 7    \hspace{0cm}\\
   \text{Strongly Disagree} \hspace{0.1cm} \(\bigcirc\) \hspace{0.05cm} \(\bigcirc\) \hspace{0.05cm} \(\bigcirc\) \hspace{0.05cm} \(\bigcirc\) \hspace{0.05cm} \(\bigcirc\) \hspace{0.05cm} \(\bigcirc\) \hspace{0.05cm} \(\bigcirc\)  \hspace{0.1cm} \text{Strongly Agree} \\
\end{center}

    \vspace{3px}

\paragraph{\textbf{Validation: }Please rate the following statement: ``The feedback provided by the \systemname system made me feel more confident and comfortable about my design.''}
\begin{center}
     \hspace{0.3cm} 1 \hspace{0.2cm} 2 \hspace{0.2cm} 3 \hspace{0.2cm} 4 \hspace{0.2cm} 5 \hspace{0.2cm} 6 \hspace{0.2cm} 7    \hspace{0cm}\\
   \text{Strongly Disagree} \hspace{0.1cm} \(\bigcirc\) \hspace{0.05cm} \(\bigcirc\) \hspace{0.05cm} \(\bigcirc\) \hspace{0.05cm} \(\bigcirc\) \hspace{0.05cm} \(\bigcirc\) \hspace{0.05cm} \(\bigcirc\) \hspace{0.05cm} \(\bigcirc\)  \hspace{0.1cm} \text{Strongly Agree} \\
\end{center}

    \vspace{3px}

\paragraph{\textbf{Ease:} Please rate the following statement: ``The \systemname system is approachable and easy to use.''}
\begin{center}
     \hspace{0.3cm} 1 \hspace{0.2cm} 2 \hspace{0.2cm} 3 \hspace{0.2cm} 4 \hspace{0.2cm} 5 \hspace{0.2cm} 6 \hspace{0.2cm} 7    \hspace{0cm}\\
   \text{Strongly Disagree} \hspace{0.1cm} \(\bigcirc\) \hspace{0.05cm} \(\bigcirc\) \hspace{0.05cm} \(\bigcirc\) \hspace{0.05cm} \(\bigcirc\) \hspace{0.05cm} \(\bigcirc\) \hspace{0.05cm} \(\bigcirc\) \hspace{0.05cm} \(\bigcirc\)  \hspace{0.1cm} \text{Strongly Agree} \\
\end{center}

    \vspace{3px}

\paragraph{\textbf{Help With My Ability:} Please rate the following statement: ``I generated feedback I would have otherwise not been able to gather on my own.''}
\begin{center}
     \hspace{0.25cm} 1 \hspace{0.2cm} 2 \hspace{0.2cm} 3 \hspace{0.2cm} 4 \hspace{0.2cm} 5 \hspace{0.2cm} 6 \hspace{0.2cm} 7    \hspace{0cm}\\
   \text{Strongly Disagree} \hspace{0.1cm} \(\bigcirc\) \hspace{0.05cm} \(\bigcirc\) \hspace{0.05cm} \(\bigcirc\) \hspace{0.05cm} \(\bigcirc\) \hspace{0.05cm} \(\bigcirc\) \hspace{0.05cm} \(\bigcirc\) \hspace{0.05cm} \(\bigcirc\)  \hspace{0.1cm} \text{Strongly Agree} \\
\end{center}

    \vspace{3px}

\paragraph{\textbf{Effort / Reward Tradeoff:} Please rate the following statement: ``The feedback that I was able to produce was worth the effort required to produce it.''}
\begin{center}
     \hspace{0.25cm} 1 \hspace{0.2cm} 2 \hspace{0.2cm} 3 \hspace{0.2cm} 4 \hspace{0.2cm} 5 \hspace{0.2cm} 6 \hspace{0.2cm} 7    \hspace{0cm}\\
   \text{Strongly Disagree} \hspace{0.1cm} \(\bigcirc\) \hspace{0.05cm} \(\bigcirc\) \hspace{0.05cm} \(\bigcirc\) \hspace{0.05cm} \(\bigcirc\) \hspace{0.05cm} \(\bigcirc\) \hspace{0.05cm} \(\bigcirc\) \hspace{0.05cm} \(\bigcirc\)  \hspace{0.1cm} \text{Strongly Agree} \\
\end{center}

    \vspace{3px}

\paragraph{\textbf{Collaboration:} How collaborative did you feel your interaction with the \systemname system was during the design process?}
% \begin{center}
%      \hspace{0.15cm} 1 \hspace{0.2cm} 2 \hspace{0.2cm} 3 \hspace{0.2cm} 4 \hspace{0.2cm} 5 \hspace{0.2cm} 6 \hspace{0.2cm} 7    \hspace{0cm}\\
%    \text{Very Not Collaborative} \hspace{0.0cm} \(\bigcirc\) \hspace{0.0cm} \(\bigcirc\) \hspace{0.0cm} \(\bigcirc\) \hspace{0.0cm} \(\bigcirc\) \hspace{0.0cm} \(\bigcirc\) \hspace{0.0cm} \(\bigcirc\) \hspace{0.0cm} \(\bigcirc\)  \hspace{0.0cm} \text{Very Collaborative} \\
% \end{center}
\begin{center}
     \hspace{0.4cm} 1 \hspace{0.07cm} 2 \hspace{0.07cm} 3 \hspace{0.07cm} 4 \hspace{0.07cm} 5 \hspace{0.07cm} 6 \hspace{0.07cm} 7    \hspace{0cm}\\
   \text{Very Not Collaborative} \(\bigcirc\) \(\bigcirc\) \(\bigcirc\) \(\bigcirc\) \(\bigcirc\) \(\bigcirc\) \(\bigcirc\) \text{Very Collaborative} \\
\end{center}

    \vspace{3px}

\paragraph{\textbf{Ambience - Hotkeys: }How smooth and seamless is your interaction when using the \textbf{hotkeys (command + R or command + E) }in the \systemname system? }
\begin{center}
     \hspace{0.45cm} 1 \hspace{0.2cm} 2 \hspace{0.2cm} 3 \hspace{0.2cm} 4 \hspace{0.2cm} 5 \hspace{0.2cm} 6 \hspace{0.2cm} 7    \hspace{0cm}\\
   \text{Very Not Ambient} \hspace{0.1cm} \(\bigcirc\) \hspace{0.05cm} \(\bigcirc\) \hspace{0.05cm} \(\bigcirc\) \hspace{0.05cm} \(\bigcirc\) \hspace{0.05cm} \(\bigcirc\) \hspace{0.05cm} \(\bigcirc\) \hspace{0.05cm} \(\bigcirc\)  \hspace{0.1cm} \text{Very Ambient} \\
\end{center}

    \vspace{3px}

\paragraph{\textbf{Ambience - Minimal Display:} How smooth and seamless is your interaction when using the \textbf{floating voice window and hidden control panel} in the \systemname system?}
\begin{center}
     \hspace{0.45cm} 1 \hspace{0.2cm} 2 \hspace{0.2cm} 3 \hspace{0.2cm} 4 \hspace{0.2cm} 5 \hspace{0.2cm} 6 \hspace{0.2cm} 7    \hspace{0cm}\\
   \text{Very Not Ambient} \hspace{0.1cm} \(\bigcirc\) \hspace{0.05cm} \(\bigcirc\) \hspace{0.05cm} \(\bigcirc\) \hspace{0.05cm} \(\bigcirc\) \hspace{0.05cm} \(\bigcirc\) \hspace{0.05cm} \(\bigcirc\) \hspace{0.05cm} \(\bigcirc\)  \hspace{0.1cm} \text{Very Ambient} \\
\end{center}

    \vspace{3px}

\paragraph{\textbf{Disturbance - Voice:} How disruptive do you find the focus of \textbf{audio interaction} during the process?}
\begin{center}
     \hspace{0.45cm} 1 \hspace{0.2cm} 2 \hspace{0.2cm} 3 \hspace{0.2cm} 4 \hspace{0.2cm} 5 \hspace{0.2cm} 6 \hspace{0.2cm} 7    \hspace{0cm}\\
   \text{Very Not Disturbing} \hspace{0.1cm} \(\bigcirc\) \hspace{0.05cm} \(\bigcirc\) \hspace{0.05cm} \(\bigcirc\) \hspace{0.05cm} \(\bigcirc\) \hspace{0.05cm} \(\bigcirc\) \hspace{0.05cm} \(\bigcirc\) \hspace{0.05cm} \(\bigcirc\)  \hspace{0.1cm} \text{Very Disturbing} \\
\end{center}

    \vspace{3px}

\paragraph{\textbf{Disturbance - Emojis:} How disruptive do you find the \textbf{background emoji interaction} during the process?}
\begin{center}
     \hspace{0.45cm} 1 \hspace{0.2cm} 2 \hspace{0.2cm} 3 \hspace{0.2cm} 4 \hspace{0.2cm} 5 \hspace{0.2cm} 6 \hspace{0.2cm} 7    \hspace{0cm}\\
   \text{Very Not Disturbing} \hspace{0.1cm} \(\bigcirc\) \hspace{0.05cm} \(\bigcirc\) \hspace{0.05cm} \(\bigcirc\) \hspace{0.05cm} \(\bigcirc\) \hspace{0.05cm} \(\bigcirc\) \hspace{0.05cm} \(\bigcirc\) \hspace{0.05cm} \(\bigcirc\)  \hspace{0.1cm} \text{Very Disturbing} \\
\end{center}

    \vspace{3px}

\paragraph{\textbf{Control:} Please rate the following statement: ``I felt like I had control over the generated feedback.''}
\begin{center}
     \hspace{0.25cm} 1 \hspace{0.2cm} 2 \hspace{0.2cm} 3 \hspace{0.2cm} 4 \hspace{0.2cm} 5 \hspace{0.2cm} 6 \hspace{0.2cm} 7    \hspace{0cm}\\
   \text{Strongly Disagree} \hspace{0.1cm} \(\bigcirc\) \hspace{0.05cm} \(\bigcirc\) \hspace{0.05cm} \(\bigcirc\) \hspace{0.05cm} \(\bigcirc\) \hspace{0.05cm} \(\bigcirc\) \hspace{0.05cm} \(\bigcirc\) \hspace{0.05cm} \(\bigcirc\)  \hspace{0.1cm} \text{Strongly Agree} \\
\end{center}

    \vspace{3px}

\paragraph{\textbf{Trust:} Please rate the following statement: ``I trust all the feedback generated by the \systemname system.''}
\begin{center}
     \hspace{0.25cm} 1 \hspace{0.2cm} 2 \hspace{0.2cm} 3 \hspace{0.2cm} 4 \hspace{0.2cm} 5 \hspace{0.2cm} 6 \hspace{0.2cm} 7    \hspace{0cm}\\
   \text{Strongly Disagree} \hspace{0.1cm} \(\bigcirc\) \hspace{0.05cm} \(\bigcirc\) \hspace{0.05cm} \(\bigcirc\) \hspace{0.05cm} \(\bigcirc\) \hspace{0.05cm} \(\bigcirc\) \hspace{0.05cm} \(\bigcirc\) \hspace{0.05cm} \(\bigcirc\)  \hspace{0.1cm} \text{Strongly Agree} \\
\end{center}

    \vspace{3px}

\paragraph{\textbf{Age}} \_\_\_\_\_\_\_\_\_\_\_\_\_

    \vspace{3px}

\paragraph{\textbf{What is your gender identity?} Select all that apply.}
\begin{itemize}[label={},leftmargin=10px]
    \item $\Box$ Man
    \item $\Box$ Woman
    \item $\Box$ Non-binary
    \item $\Box$ Other \_\_\_\_\_\_\_\_\_\_\_\_\_
\end{itemize}

    \vspace{3px}

\paragraph{\textbf{Self-assessed proficiency level in 3D CAD design}}
\begin{itemize}[label={},leftmargin=10px]
    \item \( \bigcirc \) Entry-level / Junior
    \item  \( \bigcirc \) Mid-level
    \item  \( \bigcirc \) Experienced / Senior
    \item  \( \bigcirc \) Other \_\_\_\_\_\_\_\_\_\_\_\_\_
\end{itemize}
    \vspace{3px}

\paragraph{Here are some examples of generative AI (tools). \textbf{Please check all that you have used AT LEAST ONCE in the past month, for any purpose.} Select all that apply.}
\begin{itemize}[label={},leftmargin=10px]
    \item $\Box$ Language models (ChatGPT, Gemini, Claude etc)
    \item $\Box$ Text-to-image models or Image-to-text models (DALL-E, Midjourney, Adobe Firefly etc)
    \item $\Box$ Text-to-audio/speech models or audio-to-text models
    \item $\Box$ Text-to-video models or video-to-text models 
    \item $\Box$ Other \_\_\_\_\_\_\_\_\_\_\_\_\_
\end{itemize}

    \vspace{3px}
    \paragraph{\textbf{How frequently do you use generative AI?}}
\begin{itemize}[label={},leftmargin=10px]
    \item \( \bigcirc \) Daily (at least 1 time per day)
    \item  \( \bigcirc \) Weekly (1-5 times per week)
    \item  \( \bigcirc \) Monthly (1-5 times per month)
    \item  \( \bigcirc \) Less than once every month
    \item  \( \bigcirc \) Never
    \item  \( \bigcirc \) Other\_\_\_\_\_\_\_\_\_\_\_\_\_
\end{itemize}

    \vspace{3px}
\paragraph{\textbf{How do you describe your understanding of generative AI?}}
\begin{itemize}[label={},leftmargin=10px]
    \item \( \bigcirc \) Substantial understanding---I have in-depth knowledge of generative AI models and tools.
    \item  \( \bigcirc \) Good general understanding---I’ve read about generative AI and grasp the key concepts.
    \item  \( \bigcirc \) Basic understanding---I have vague idea of generative AI.
    \item  \( \bigcirc \) Minimal understanding---I know litte about generative AI.
    \item  \( \bigcirc \) Other\_\_\_\_\_\_\_\_\_\_\_\_\_
\end{itemize}

    \vspace{3px}
\paragraph{\textbf{Before today's study, have you ever used generative AI tools in your design process?} Select all that apply.}
\begin{itemize}[label={},leftmargin=10px]
    \item $\Box$ for ideation or brainstorming.
    \item $\Box$ for creating prototypes and generating design variations.
    \item $\Box$ for enhancing or refining existing designs.
    \item $\Box$ for automating repetitive tasks.
    \item $\Box$ No, I have not used generative AI tools in my design process.
    \item $\Box$ Other \_\_\_\_\_\_\_\_\_\_\_\_\_\_\_\_\_\_\_\_\_\_\_\_\_\_
\end{itemize}

    \vspace{3px}
\paragraph{\textbf{Perceived Stakeness/Pressure - AI Feedback:} How much pressure do you feel when receiving \textbf{AI feedback}  for your 3D design process today? (1 = Very Low Stakes: I don’t feel stressed; 7 = Very High Stakes: I feel a lot of pressure.}
\begin{center}
     \hspace{0cm} 1 \hspace{0.2cm} 2 \hspace{0.2cm} 3 \hspace{0.2cm} 4 \hspace{0.2cm} 5 \hspace{0.2cm} 6 \hspace{0.2cm} 7    \hspace{0.7cm}\\
   \text{Very Low Stakes} \hspace{0.1cm} \(\bigcirc\) \hspace{0.05cm} \(\bigcirc\) \hspace{0.05cm} \(\bigcirc\) \hspace{0.05cm} \(\bigcirc\) \hspace{0.05cm} \(\bigcirc\) \hspace{0.05cm} \(\bigcirc\) \hspace{0.05cm} \(\bigcirc\)  \hspace{0.1cm} \text{Very High Stakes} \\
\end{center}

    \vspace{3px}
\paragraph{\textbf{Perceived Stakeness/Pressure - Human Feedback: }How much pressure do you feel when receiving \textbf{human feedback} for your 3D design process? (1 = Very low-stakes: I don’t feel stressed; 7 = Very high-stakes: I feel a lot of pressure.}
\begin{center}
     \hspace{0cm} 1 \hspace{0.2cm} 2 \hspace{0.2cm} 3 \hspace{0.2cm} 4 \hspace{0.2cm} 5 \hspace{0.2cm} 6 \hspace{0.2cm} 7    \hspace{0.7cm}\\
   \text{Very Low Stakes} \hspace{0.1cm} \(\bigcirc\) \hspace{0.05cm} \(\bigcirc\) \hspace{0.05cm} \(\bigcirc\) \hspace{0.05cm} \(\bigcirc\) \hspace{0.05cm} \(\bigcirc\) \hspace{0.05cm} \(\bigcirc\) \hspace{0.05cm} \(\bigcirc\)  \hspace{0.2cm} \text{Very High Stakes} \\
\end{center}

    \vspace{3px}
\paragraph{\textbf{Perceived Pressure - Actions:} How does the perceived pressure from seeking feedback from AI and humans affect your future actions?}
\begin{itemize}[label={},leftmargin=10px]
    \item \( \bigcirc \) I prefer AI feedback because it feels less stressful and more accessible.
    \item  \( \bigcirc \) I prefer human feedback because it feels more critical and valuable.
    \item  \( \bigcirc \) I find that perceived pressure does not significantly impact my choice between AI and human feedback.
    \item  \( \bigcirc \) Other\_\_\_\_\_\_\_\_\_\_\_\_\_
\end{itemize}

    \vspace{3px}

\paragraph{\textbf{Attitude Towards AI for Design}: How would you describe your attitude towards integrating generative AI into your design process?}
\begin{itemize}[label={},leftmargin=10px]
    \item \( \bigcirc \) Very Positive
    \item  \( \bigcirc \) Positive
    \item  \( \bigcirc \) Neutral
    \item \( \bigcirc \)  Negative
    \item  \( \bigcirc \) Very Negative
    \item  \( \bigcirc \) Other\_\_\_\_\_\_\_\_\_\_\_\_\_
\end{itemize}

% % \includepdf[pages=-,scale=1]{fig/survey.pdf}
% \begin{figure}[!h]
% \vspace{-20px}
% \includegraphics[width=.96\linewidth,page=1]{fig/survey.pdf}
% \vspace{-10px}
% \end{figure}

% \begin{figure}[!h]
% \includegraphics[width=.96\linewidth,page=2]{fig/survey.pdf}
% \end{figure}

% \begin{figure}[!h]
% \includegraphics[width=.96\linewidth,page=3]{fig/survey.pdf}
% \end{figure}

% \begin{figure}[!h]
% \includegraphics[width=.96\linewidth,page=4]{fig/survey.pdf}
% \end{figure}

% \newpage
\section{Post-Study Interview Questions:} 

    \vspace{3px}

\paragraph{\textbf{Numbers of the Feedback}}
You have requested \underline{\hspace{0.5cm}} times of feedback during the design process. However, if you were to follow the original workflow, would you request this amount of feedback from ChatGPT or a friend? Why?

    \vspace{3px}

\paragraph{\textbf{Pros}}
What benefits did you experience after using our tool? Did it help with confidence, validation, emotional support, ease of getting quick feedback, or offering diverse perspectives?

    \vspace{3px}

\paragraph{\textbf{Cons}}
What are some shortcomings you find after using our tool? If we ask you to redesign the system, what would you add or remove from the tool?
    \vspace{3px}

\paragraph{\textbf{Human VS AI}}
Compared to human feedback, what’s new with AI feedback? Do you prefer AI feedback? Do you see any opportunities or challenges with it?

    \vspace{3px}

\paragraph{\textbf{Quality of the Feedback -Usefulness:}} How useful do you feel about the overall feedback?

    \vspace{3px}

\paragraph{\textbf{Quality of the Feedback - Relevance:}} How relevant do you feel the overall feedback was? Did you find the system understood your design context and provided accurate feedback based on it?

    \vspace{3px}

\paragraph{\textbf{Interaction of the Feedback}}
How do you like the idea of ambient design? For instance, requesting feedback using hotkeys and receiving feedback via audio interaction.

    \vspace{3px}

\paragraph{\textbf{Future Usage:}} If you had this tool available in future design projects, would you use it? Would the usage be similar or would you have new use cases in real-time?

    \vspace{3px}

\paragraph{\textbf{Workflow Integration:}} How might this tool impact your current design workflow?

% \end{document}

\end{document}